\documentclass[twocolumn,twocolappendix]{aastex631}
\pdfoutput=1 
\usepackage{amsmath,amstext}
\usepackage[T1]{fontenc}
\usepackage[figure,figure*]{hypcap}
\usepackage{mwe}
\usepackage{lipsum}
\usepackage{graphicx} 
\usepackage{gensymb}
\usepackage{multirow}
\usepackage{xcolor,colortbl}
\usepackage{tipa}
\usepackage{natbib}
\usepackage{hyperref}
\usepackage{ulem}
\usepackage{float}
\usepackage{xfrac}

\newcommand{\gaia}{\textit{Gaia}\xspace}

\newcommand{\LB}{Local Bubble\xspace}
\newcommand{\LC}{Local Chimney\xspace}
\newcommand{\lmax}{$l_{\rm{max}}$\xspace}
\newcommand{\nside}[1]{$N_{\rm{side}} = {#1}$\xspace}
\newcommand\HI{$\textrm{H}\scriptstyle\mathrm{I}$}

\newcommand{\ellb}[2]{$(\ell, b) = ({#1}\degree, {#2}\degree)$}

\newcommand{\perc}[1]{P$_{#1}$\xspace}
\newcommand{\Pmin}{$P_{\rm{min}}$\xspace}
\newcommand{\Aprime}{$A'_{ZGR23}$\xspace}
\newcommand{\Npeak}{$N_{peak}$\xspace}
\newcommand{\GSHbub}{GSH238+00+09\xspace}
\newcommand{\Hpx}{HEALPix\xspace}
\newcommand{\dinner}{$d_{inner}$\xspace}
\newcommand{\douter}{$d_{outer}$\xspace}

\shorttitle{Local Bubble is a Local Chimney}
\shortauthors{O'Neill et al.}

\begin{document}

\title{The Local Bubble is a Local Chimney: A New Model from 3D Dust Mapping}

\author[0000-0003-4852-6485]{Theo J. O'Neill}
\affiliation{Center for Astrophysics $|$ Harvard \& Smithsonian, 60 Garden St., Cambridge, MA 02138, USA}

\author[0000-0002-2250-730X]{Catherine Zucker}
\affiliation{Center for Astrophysics $|$ Harvard \& Smithsonian, 60 Garden St., Cambridge, MA 02138, USA}

\author[0000-0003-1312-0477]{Alyssa A. Goodman}
\affiliation{Center for Astrophysics $|$ Harvard \& Smithsonian, 60 Garden St., Cambridge, MA 02138, USA}

\author[0000-0003-3122-4894]{Gordian Edenhofer}
\affiliation{Max Planck Institute for Astrophysics, Karl-Schwarzschild-Straße 1, 85748 Garching, Germany}
\affiliation{Ludwig Maximilian University of Munich, Geschwister-Scholl-Platz 1, 80539 Munich, Germany}

\begin{abstract}
Leveraging a high-resolution 3D dust map of the solar neighborhood from Edenhofer et al. (2024), we derive a new 3D model for the dust-traced surface of the Local Bubble, the supernova-driven cavity surrounding the Sun.  We find that the surface of the Local Bubble is highly irregular in shape, with its peak extinction surface falling at an average distance of 170 pc from the Sun (spanning 70--600+ pc) with a typical thickness of 35 pc and a total dust-traced mass of $(6.0 \pm 0.7) \times 10^5 \ \rm{M}_{\odot}$.  The Local Bubble displays an extension in the Galactic Northern hemisphere that is morphologically consistent with representing a ``Local Chimney.'' We argue this chimney was likely created by the ``bursting'' of this supernova-driven superbubble, leading to the funneling of interstellar medium ejecta into the lower Galactic halo.  We find that many well-known dust features and molecular clouds fall on the surface of the Local Bubble and that several tunnels to other adjacent cavities in the interstellar medium may be present.  Our new, parsec-resolution view of the Local Bubble may be used to inform future analysis of the evolution of nearby gas and young stars, the investigation of direct links between the solar neighborhood and the Milky Way's lower halo, and numerous other applications.  
\end{abstract}

\section{Introduction}\label{S:intro}

Over the last five decades, a picture has emerged of the Sun residing in a low density interstellar cavity now known as the \LB \citep[see reviews in][]{CoxReynolds1987, WelshShelton2009, LinskyRedfield2021}.  Spanning a few hundred parsecs in diameter, a variety of evidence suggests the \LB is a supernova-driven superbubble, where sequential supernovae drove the creation of an evacuated interior cavity surrounded by a shell of swept-up dust and gas   --- with indicators including recent star formation on timescales consistent with supernova-driven shell expansion \citep[e.g.,][]{FuchsBreitschwerdt2006,ZuckerGoodman2022}, a correlation in the Earth's geological record between the $^{60}$Fe isotope and the timing of the supernovae that shaped the Bubble \citep[e.g.,][]{BenitezMaiz-Apellaniz2002,BreitschwerdtFeige2016,SchulreichFeige2023}, and the possible presence of warm-to-hot X-ray emitting plasma in the Bubble's interior (e.g., \citealt{SandersKraushaar1977,SnowdenEgger1998}, although some fraction of this X-ray emission stems from Solar wind charge exchange (SWCX) \citep{WelshShelton2009}). 

Many tracers have been used to generate 3D maps of the \LB, including Na$\mathrm{I}$ absorption measurements \citep{SfeirLallement1999,LallementWelsh2003}, stellar color excess measurements \citep{LallementVergely2014}, X-ray emission \citep{SnowdenEgger1998,LiuChiao2017}, and diffuse interstellar bands \citep{FarhangvanLoon2019}.  Thanks to the advent of 3D dust mapping of the solar neighborhood within the last decade  \citep[e.g.,][]{GreenSchlafly2015,LeikeEnsslin2019}, reconstructing the \LB's shell as a region of higher dust density is now possible.  \citet{PelgrimsFerriere2020} mapped the geometry of the Local Bubble's shell using the 25-pc-resolution 3D dust map of \citet{LallementBabusiaux2019}; their model enabled detailed analysis of the Bubble's relationship to nearby molecular clouds and star-forming regions by \citet{ZuckerGoodman2022}, who found that nearly all recent star formation within 200 pc of the Sun was triggered by the Bubble's supernova-driven expansion over the last $\sim$14 Myr.  

There are many open questions about the specifics of the \LB's morphology, including whether the \LB is closed or open at high latitudes, i.e., if the ``bubble'' is actually a ``chimney'' that has broken out of the Galactic disk and is funneling material into the Milky Way's halo.  Burst bubbles and Galactic chimneys are common in simulations and theoretical predictions \citep[e.g.,][]{MacLowMcCray1989, deAvillezBerry2001, FieldingQuataert2018,OrrFielding2022_turbulence}, where the interstellar medium (ISM) is enriched via a Galactic fountain flow \citep{ShapiroField1976} in which sufficiently energetic superbubbles can break out of the dense gas in the Galactic plane and form chimneys that vent energy and enriched ISM material into the halo \citep{NormanIkeuchi1989}, some of which may then fall back to the disk as intermediate velocity clouds \citep{Bregman1980,Kahn1981,Kahn1991,KahnBrett1993}.  

Early maps supported a picture of the \LB as a \LC \citep[e.g.,][]{SfeirLallement1999, WelshSfeir1999, VergelyFreireFerrero2001, LallementWelsh2003}, but more recent dust-based models of the \LB \citep{PelgrimsFerriere2020} have presented a view of the \LB as a closed surface.  Constraining the high-altitude dust-traced surface of the \LB (and the relationship between the \LB and the local Galactic halo) requires a high-resolution view of low density dust.  The recent 3D dust map from \citet{EdenhoferZucker2024} provides a parsec-scale, all-sky view of dust over a large dynamic range and out to distances of 1.25 kpc from the Sun --- ideal for reconstructing the geometry of the \LB.

In this work, we map the 3D shell of the \LB using the \citet{EdenhoferZucker2024} dust map.  We find that the \LB has morphological features consistent with being an asymmetric \LC with an open Northern cap.  In \S\ref{S:data} we describe the methods we use to model the shape of the \LB using the \citet{EdenhoferZucker2024} map.  We summarize the derived properties of the \LB's shell in \S\ref{S:results}.  In \S\ref{S:discuss} we discuss the morphological features of our model, including the nature of the Chimney feature, potential ``tunnels'' to adjacent bubbles, the positions of molecular clouds and prominent dust features relative to the \LB's surface, and the significance of the \LC in the context of a Milky Way whose ISM and lower halo are influenced by feedback-driven bubbles.  We conclude in \S\ref{S:conclude}.

\section{Data and Methods}\label{S:data}

\subsection{Edenhofer et al. (2024) 3D Dust Map}

We use the \citet[][hereafter E24]{EdenhoferZucker2024} 3D map of dust within 1.25 kpc of the Sun to build a new model of the \LB's surface. The E24 map is capable of probing to lower densities at higher altitudes off the Galactic plane than previous maps at comparable resolution \citep[see e.g., the similarly spatially resolved map of][]{LeikeGlatzle2020}.  E24 modeled the logarithm of the 3D dust extinction density using a Gaussian Process (implemented in {\tt NIFTy.re}, \citealt{EdenhoferFrank2024})  and a new technique known as Iterative Charted Refinement \citep{EdenhoferLeike2022}, which imposes a correlation kernel over arbitrarily-spaced voxels, iteratively refining the resolution of the map (from coarse to fine) until it achieves the desired resolution. The E24 map was constructed using the \citet[][herafter ZGR23]{ZhangGreen2023} stellar distance and extinction estimates derived from Gaia BP/RP spectra.  

The E24 map is defined in unitless extinction ZGR23, with sampled differential extinction measurements given by
\begin{equation}
    A'_{ZGR23} =  \frac{d A_{ZGR23}}{1 \textrm{pc}}.
\end{equation}
Our surface finding method is agnostic to the exact wavelength of extinction probed, but this measurement can be converted to other bands using ZGR23's published extinction curve (as described in \S\ref{S:extinc_dens_mass}).

The map was constructed in a spherical coordinate system with logarithmically spaced distances, with an angular spacing of \Hpx \citep{GorskiHivon2005} \nside{256} (equivalent to 13.7' pixel size) and pc-scale distance resolution, sampling the dust in distance bins ranging in size between 0.4--7 pc.  The E24 map does not include the region within a distance $d < 69$ pc of the Sun; our model of the \LB is therefore not sensitive to this nearby volume, which includes the low-density complex of clouds known as the ``Local Fluff'' in which the Sun is directly immersed \citep{Frisch1986}.  The center of the maximum distance bin included in the map falls at $d=1244$ pc.

In generating their map, E24 drew 12 samples from their inferred distribution of 3D dust extinction.  We derive our model of the \LB from the posterior mean of their reconstruction (i.e., the mean of the 12 samples), while using the 12 samples to constrain the statistical uncertainty (see \S\ref{S:uncert_prop} and Appendix \ref{ap:uncertainty}). We additionally make available the individual models of the \LB derived from each of the 12 samples.

We query the mean E24 map along each line-of-sight (LOS) from the Sun in Galactic spherical coordinates ($\ell$, $b$, $d$)\footnote{We later convert this to a heliocentric Cartesian coordinate system ($x$, $y$, $z$) pc, where $x$ points from the Sun to the Galactic center at Galactic longitude $\ell$ = 0$\degree$, $y$ is oriented towards $\ell$ = 90$\degree$, and $z$ is oriented towards the North Galactic Pole at Galactic latitude $b$ = 90$\degree$,
\begin{equation}
\nonumber
\begin{split}
   x &= d \cos (\ell) \cos(b) \\
   y &= d \sin(\ell) \cos(b) \\
   z &= d \sin(b).
\end{split}
\end{equation}
}
using the python package {\tt dustmaps} \citep{M_Green_2018}, with LOS spaced as \nside{256}.  For each LOS, we sample the E24 dust map between 69 pc -- 1244 pc at uniform $dr = 1$ pc intervals.  

\subsection{Peak Finding Method}\label{S:peak_find}

\begin{figure*}
    \centering
    \includegraphics[width=\textwidth]{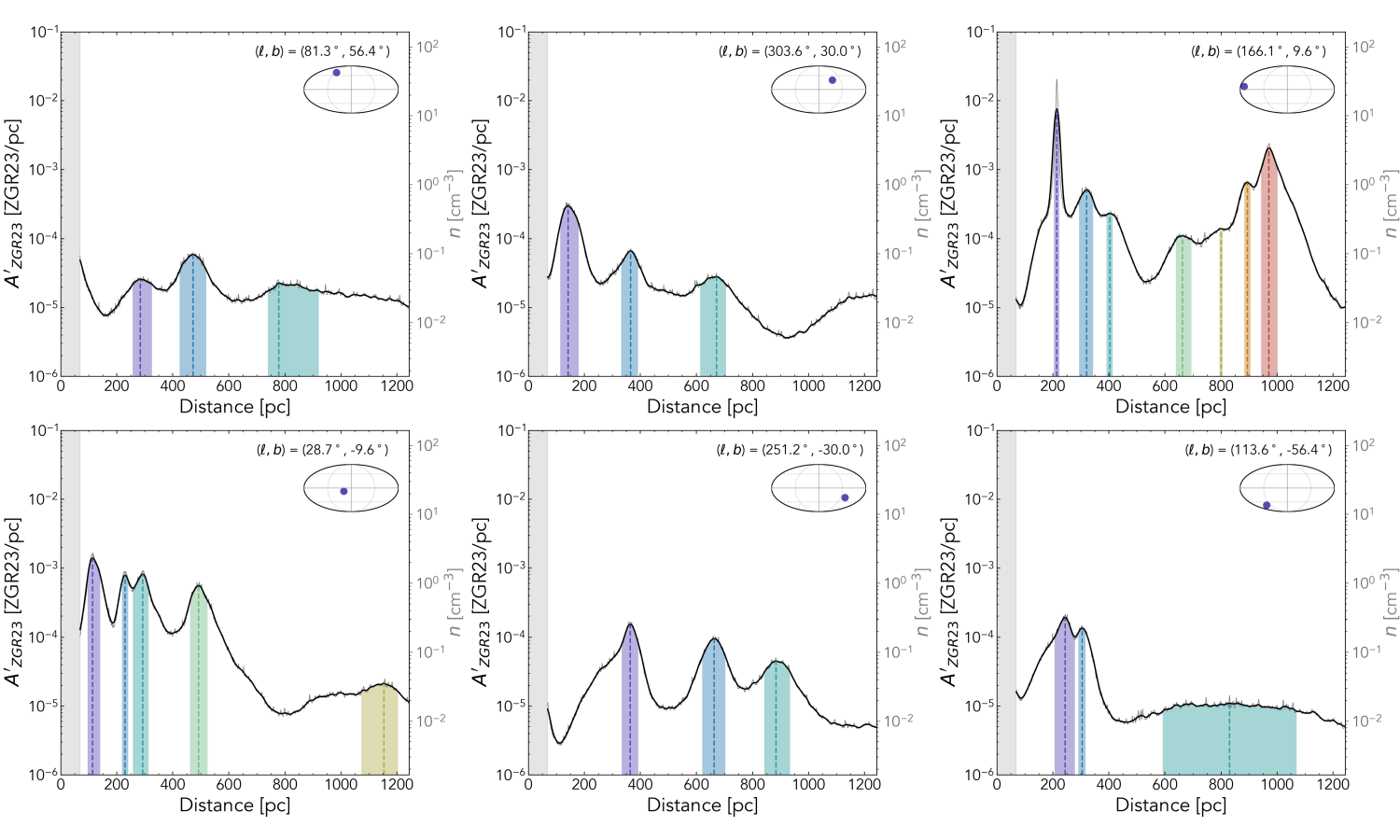}
    \caption{Differential dust extinction in the E24 map (\Aprime) as a function of distance, along six representative lines of sight. The raw E24 extinction is shown in gray and the smoothed extinction is shown in black.  Peaks identified in the dust map are shown by the colored shaded regions, with peak distance $d_{peak}$ marked by a dashed line and peak shading extending from \dinner to \douter.  The first peak along the LOS (corresponding to the shell of the \LB) is always shaded in purple.  The inner $d<69$ pc region where the E24 map is undefined is shaded in gray.  The secondary y-axis shows the conversion from \Aprime to volume density $n$. The inset Mollweide projection shows the position of each LOS on-sky.
    }
    \label{fig:peakfind}
\end{figure*}

We define the \LB's shell as the first significant peak in extinction along the LOS from the Sun.\footnote{\citet{PelgrimsFerriere2020} also define their map of the \LB's shell as the first peak along the LOS.  See Appendix \ref{ap:pelgrims} for a discussion of the difference in peak finding methods used.}  We define a peak as local maximum along the LOS with a prominence greater than some given value (where ``prominence'' is the height of the peak above its base) .  Our technique is predicated on the idea that the vacuous inner cavity of the Local Bubble is bounded by a denser shell of neutral gas and dust; however, as we will discuss in \S \ref{S:extinc_dens_mass}, our methodology is insensitive to the precise density of the shell.

We smooth the LOS extinction profile using a Gaussian kernel with $\sigma_{smooth} = 7$ pc to reduce the effects of noise in our peak detection, and require a minimum peak prominence of $P_{min} = 2.04 \times 10^{-6}$ ZGR23/pc.  We describe the selection of these parameters in Appendix \ref{ap:peakfind_params}.  We identify peaks using the python package {\tt scipy}'s {\tt find\_peaks} function. 

We define the inner and outer edges of each peak as the distance before (\dinner) and distance after (\douter) the peak at which the differential extinction \Aprime is equal to
\begin{equation}
    A'_{0.5} = A'_{peak} - 0.5 P, 
\end{equation}
i.e., the peak height minus half the peak prominence.  If the half prominence boundaries contain a separate but lower prominence peak, we adjust the inner (or outer) edge of the overlapping peak to fall at the local minimum between the two peaks.  We additionally derive and report the inner and outer edges (and all other associated properties of our \LB model) at a more generous threshold of $A'_{0.9}$.  We define the thickness of a given peak as $\Delta_{peak} =$ \douter$ - $ \dinner, and require a minimum thickness (at $A'_{0.5}$) of 3 pc so that \dinner$ \neq d_{peak} \neq$ \douter.  

Six representative LOS with identified peaks are shown in Figure \ref{fig:peakfind}.  Wide variations in peak shape, width, and height are present across the sky.  

We first performed our peak finding method on the full E24 map out to $d=1244$ pc.  For the vast majority of LOS, the first peak along the LOS was within a distance of $d=650$ pc, with only 23 (0.003\%) isolated LOS at high latitudes having peak distances beyond $d=650$ pc.  We interpret these isolated LOS as being unassociated with the \LB and, for our final model of the \LB, performed our peak finding method on the E24 map truncated to $d=650$ pc to exclude these handful of LOS well beyond the surface of the \LB.

\vspace{5em}

\subsection{Derived Peak Properties}\label{S:derive_props}

\begin{deluxetable*}{cc|ccccc|c}
\centering
\tablecaption{Properties of the Local Bubble's Shell \label{tab:shell_props}} 
\tablehead{\colhead{Property} & \colhead{Symbol} & \colhead{Minimum} & \colhead{\perc{2.28}} & \colhead{Median} & \colhead{\perc{97.72}} & \colhead{Maximum} & \colhead{Units}}
\startdata	
Inner Edge & \dinner & 69 & 80 & 150 & 322 & 592 & pc \\
Peak Distance & $d_{peak}$ & 71 & 91 & 170 & 362 & 616 & pc \\
Outer Edge & \douter & 72 & 101 & 191 & 395 & 633 & pc \\
\hline
Thickness & $\Delta_{shell}$ & 3 & 13 & 35 & 123 & 357 & pc \\
Peak Density & $n_{peak}$ & 0.02 & 0.04 & 0.61 & 21 & 770 & cm$^{-3}$ \\
Extinction & $A_G$ & <0.001 & 0.002 & 0.02 & 0.34 & 2.8 & mag \\
Inclination & $\gamma$ & 0 & 4 & 25 & 63 & 90 & $\degree$ \\
\enddata
\tablecomments{\perc{i} denotes the $i$-th percentile of the distribution.  Under a normal distribution, \perc{2.28} and \perc{97.72} represent the $\mp 2\sigma$ intervals, respectively.  See Figure \ref{fig:shell_stats} for a graphical representation of these distributions.}
\end{deluxetable*}

For each LOS, we calculate various properties of the \LB's shell, including extinction, density, mass, and inclination to the plane-of-the-sky.

\subsubsection{Extinction, Density, and Mass}\label{S:extinc_dens_mass}

For a peak along the LOS extending over $i$ distance slices between \dinner $\leq d \leq$ \douter, integrated extinction can be calculated as the sum of unsmoothed differential extinction,
\begin{equation}
    A_{ZGR23} = \sum_i A'_{ZGR23, i} dr_i,
\end{equation}
where $dr_i = 1$ pc for all slices.  ZGR23 extinction can be converted to \gaia G-band extinction $A_G$ \citep[centered at $\lambda = 673$ nm,][]{JordiGebran2010} using ZGR23's published extinction curve,
\begin{equation}
    A_G = 2.0407 A_{ZGR23}.
\end{equation}
We report the integrated $A_G$ for each LOS.  We emphasize that uncertainties (calculated as in \S\ref{S:uncert_prop}) on extinctions below $A_G \simeq 0.005$ mag are generally very high (with distance uncertainties typically $\gtrsim 50$ pc).

The total volume density of hydrogen nuclei $n$ within each of the $i$ distance slices along a given LOS can be derived from extinction by following \citet{zucker2021} in assuming the ratio of hydrogen column density to extinction is constant ($A_G / N_H \simeq 4.0 \times 10^{-22}$ cm$^{2}$ mag, \citealt{Draine2003,Draine2009}), leading to a relationship
\begin{equation}
\begin{split}
    n_i & = \frac{1}{(A_G / N_H)} \left(\frac{d A_{G,i}}{1 \textrm{pc}}\right) \frac{da_i \ dr_i}{dv_i} \\ 
    &= 1653 \ \textrm{cm}^{-3} \ A'_{ZGR23, i},
    \label{eqn:vol_dens}
\end{split}
\end{equation}
where $da_i$ is the projected physical area of the pixel in distance slice $i$, $dr_i$ is the radial separation between slice $i$ and $i+1$, and $dv_i = da_i dr_i$ is the volume spanned between slice $i$ and $i+1$.  We summarize the derivation of this relationship in Appendix \ref{ap:voldens}.  For each LOS, we report the maximum unsmoothed density, $n_{peak}$, reached between \dinner and \douter.  We additionally provide an interpolated 3D grid (in heliocentric Cartesian x-y-z space) of dust in the \LB's shell.  

When integrated over the $i$ slices between \dinner $\leq d \leq $\douter, volume density yields mass contained in the peak,
\begin{equation}
    M =  1.37 \ m_p \sum_i \ n_i  \ dv_i ,
    \label{eqn:mass}
\end{equation}
where $m_p$ is the mass of a proton and 1.37 is a factor derived from cosmic abundances to convert from hydrogen mass to total mass including helium.  To calculate $dv_i$, we approximate $da_i \simeq d_i^2 \theta^2$, where $d_i$ is the distance to slice $i$ and $\theta^2$ is the area of the \Hpx pixel (which at \nside{256} is equal to $\theta^2 \simeq 1.6 \times 10^{-5}$ rad$^{2}$).  Radial separation is equal to $dr_i = 1$ pc for all slices.  
\subsubsection{Inclination}
 
The angle of the \LB's shell to the plane-of-the-sky (POS), or the shell's inclination, may be a significant factor in observations of e.g., cosmic ray deflection, pulsar scintillation, polarization fractions, and other similar quantities measured from our vantage point in the Solar System \citep[see e.g.][]{Ocker2024}.  We estimated the inclination of the \LB's shell (at $d=d_{peak}$) for each LOS by fitting a tangent plane via singular value decomposition (SVD) to its 500 nearest neighboring peaks in 3D Cartesian space (including the central point, and with neighborhoods typically spanning a space 5--25 pc in radius).

Specifically, for each LOS and collection of neighbors we perform SVD as implemented in the python package {\tt numpy} \citep{harris2020_numpy}, where decomposition is performed as $A = U \Sigma V^T$, where $A$ is a 3$\times$500 matrix of point positions (with coordinates shifted to an origin at their mean position), $U$ is a 3$\times$3 matrix with columns $(\bf{u}_1, \bf{u}_2, \bf{u}_3)$, $V$ is a 500$\times$500 matrix with columns $(\bf{v}_1, ...,\bf{v}_{500})$, and $\Sigma$ is a diagonal matrix with the singular values of $A$ as the diagonal entries, $\sigma_i = \Sigma_{ii}$.  The normal vector to the tangent plane, $\textbf{n} = (n_{x}, n_{y}, n_{z})$, can then be extracted as the third left singular vector, $\textbf{n} = \bf{u}_3$. 

The angle between the \LB's surface and the POS can then be defined as the angle between $\textbf{n}$ and the LOS (the normal to the POS),
\begin{equation}
    \gamma = \arccos (\textbf{n} \cdot \textbf{e}_L), \rm{\ \ } \gamma \in [0\degree,90\degree],
\end{equation}
where $\textbf{e}_L$ is the normalized LOS (a unit vector from the Sun to the \LB's surface, $\textbf{e}_L = (x/d, y/d, z/d)$).

\subsubsection{Uncertainties}\label{S:uncert_prop}

We estimate statistical uncertainties on our derived peak properties by applying our peak finding method to each of the 12 draws of the E24 map (with results described in Appendix \ref{ap:uncertainty}, and Figure \ref{fig:multi_draws} showing projected 2D and interactive 3D views of the \LB model derived from each draw).  Uncertainties along each LOS are then defined as the standard deviation of the draw-derived properties.  We report uncertainties for all quantities except on-sky position and derived Cartesian coordinates.  

We estimate the uncertainty on the total mass of the \LB's shell in two parts: 1) the statistical uncertainty as the standard deviation of the total masses in each draw, and 2) the systematic uncertainty as the uncertainties introduced by A) our procedure for defining peak edges (affecting the total extinction $A_G$ contributed by the shell) and B) our conversion from extinction to mass (affecting the factor $A_G/N_H$).

We expect the bulk of the uncertainty stemming from A) is contributed by the size of the smoothing kernel applied to differential extinction along the LOS, $\sigma_{smooth}$; shell thickness increases with smoothing scale, which in turn increases shell extinction and mass estimates.  To quantify this effect, we calculated the total mass of the \LB's shell $M_{shell}(\sigma_{smooth})$ for smoothing scales between $\sigma_{smooth} = 3$ pc and $\sigma_{smooth} = 10$ pc (using the same fiducial \Pmin for all $\sigma_{smooth}$).  We performed ordinary least squares regression to estimate the slope of the relationship between smoothing kernel size and the percent difference between $M_{shell}(\sigma_{smooth})$ and the fiducial $M_{shell}(7 \ \textrm{pc})$, defined as $\delta M_{shell} = [M_{shell}(\sigma_{smooth}) - M_{shell}(7 \ \textrm{pc})]/[M_{shell}(7 \ \textrm{pc})]$.  We find that, for an increase in $\sigma_{smooth}$ of 1 pc, $\delta M_{shell}$ will on average increase by 5.8$\% \pm 0.1 \%$.  Standard diagnostics of the fit suggest this model is adequate at predicting $\delta M_{shell}$ (coefficient of determination $R^2 = 0.998$, $F$-test statistic of $F=3941$ with $p < 0.001$).  We use this factor as a proxy for the systematic uncertainty on shell mass stemming from A). 

We estimate the uncertainty from B) by using the simplifying assumption that the majority of the uncertainty is introduced by the assumption of a fixed reddening vector $R_V$ used to derive the relationship between $A_\lambda/N_H$ by \citealt{Draine2003}.  Our adopted $A_G/N_H \simeq 4.0 \times 10^{-22}$ cm$^2$ mag was calculated using the relationships derived by \citet{Draine2003} for a fixed $R_V = 3.1$ \citep{CardelliClayton1989}; if instead $R_V = 2.8$ or $R_V = 3.2$ were assumed (the range within which ZGR23's derived extinction curve is consistent with \citealt{CardelliClayton1989}), factors of $A_G/N_H \simeq 3.5 \times 10^{-22}$ cm$^2$ mag or $A_G/N_H \simeq 4.1 \times 10^{-22}$ cm$^2$ mag, respectively, would result.  We use this $\sim$10\% variation as a proxy for the uncertainty on $A_G/N_H$.

Total shell mass is proportional to the product of integrated $A_G$ (which is influenced by smoothing scale) and the conversion factor $A_G/N_H$.  We expect these quantities to vary independently, so we estimate our total systematic uncertainty on total shell mass by adding our fractional uncertainties in quadrature, leading to a systematic uncertainty on $M_{shell}$ of order 12\%.

\begin{figure*}
    \centering
    \href{https://theo-oneill.github.io/localbubble/shell_dust/}{\includegraphics[width=\textwidth]{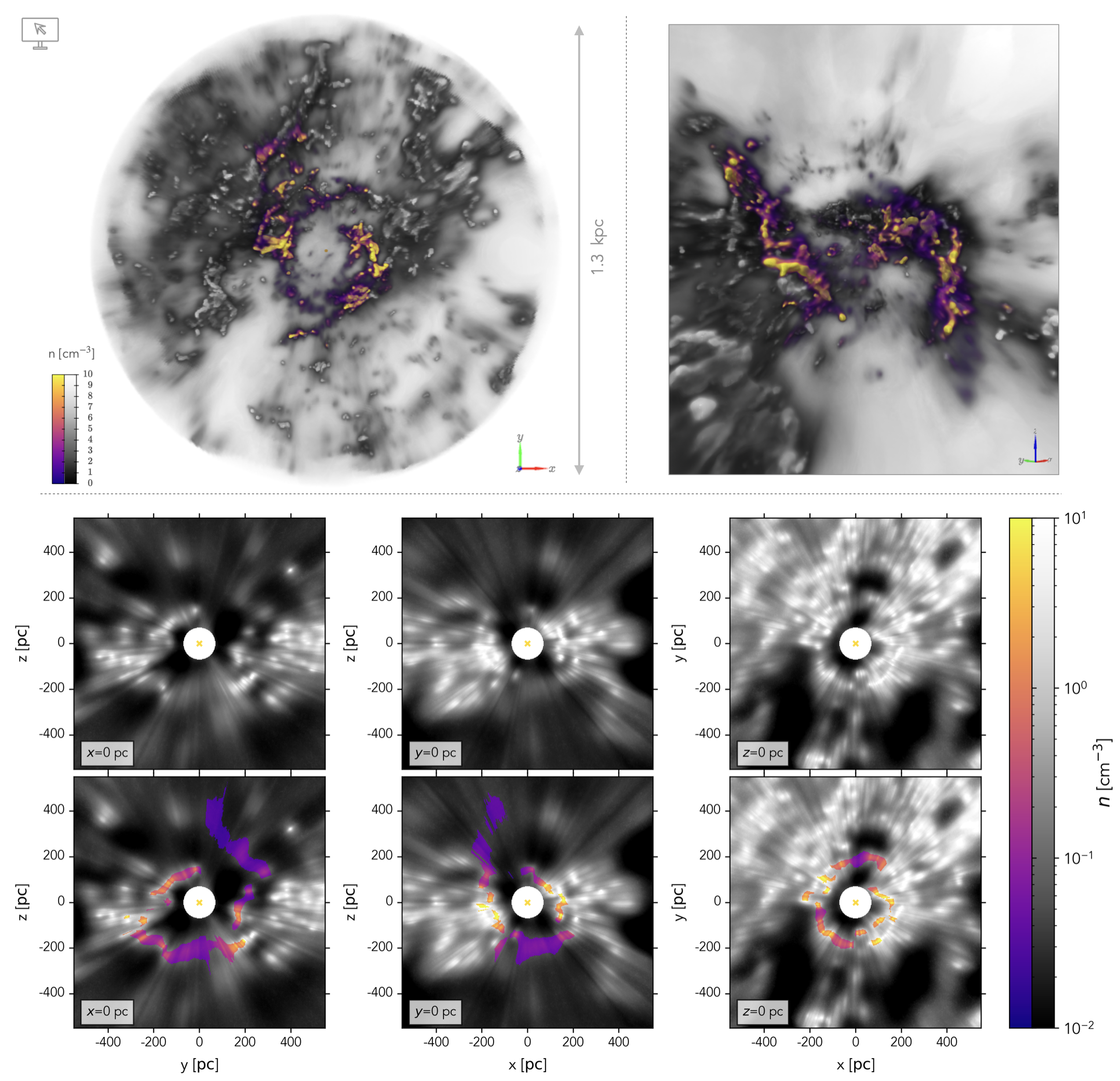}}
    \caption{Dust in the E24 map identified as being part of the \LB's shell is shown in 3D (top) and 2D projections (bottom).  \textit{Top Row:} Two views of the 3D interactive figure.  The E24 map is shown in grayscale (for $d \leq 650$ pc, with [5 pc]$^3$ voxels), and \LB shell dust is highlighted in color (with [2 pc]$^3$ voxels, smoothed with a Gaussian beam of $\sigma=2$ pc for visualization purposes).  \textit{Center \& Bottom Rows:} 2D slices through the E24 volume are shown, with slices along the $x=0$ pc (left), $y=0$ pc (center), and $z=0$ pc (right) planes.  Center row shows only the E24 map, bottom row shows the E24 map with \LB shell dust overlaid in color.  Both the full map and shell dust are shown with [2 pc]$^3$ voxels.  Color maps are scaled linearly in the 3D figure and logarithmically in the 2D figures.  \underline{Interactive 3D figure}: \url{https://theo-oneill.github.io/localbubble/shell\_dust/} }
    \label{fig:slices_2d}
\end{figure*}

\section{Results}\label{S:results}

\subsection{Properties of the Local Bubble's Shell}

\begin{figure*}
    \centering
    \href{https://theo-oneill.github.io/localbubble/distance/}{\includegraphics[width=0.97\textwidth]{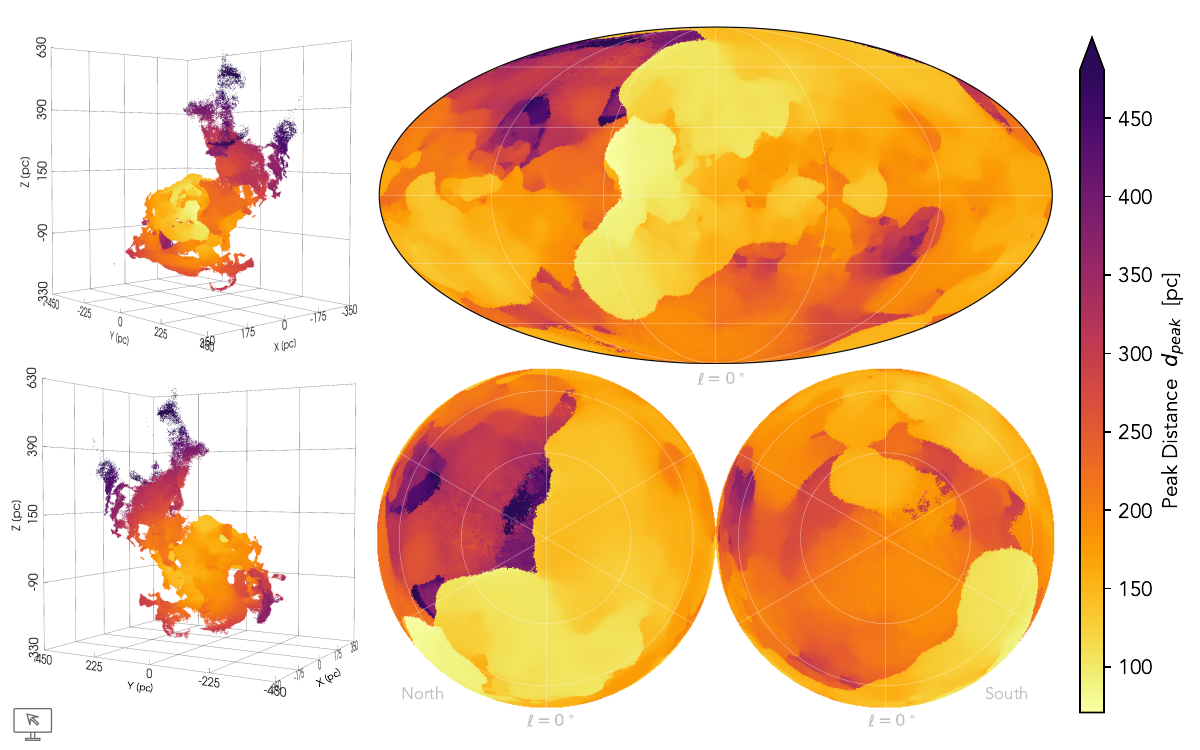}}
    \caption{Distance to the peak extinction surface of the \LB, $d_{peak}$, shown in 3D (left column) and 2D Mollweide (top right) and 2D polar (North: bottom center, South: bottom right) projections.  \underline{Interactive 3D figure} is available at: \url{https://theo-oneill.github.io/localbubble/distance/}}
    \label{fig:distance_peak}

    \href{https://theo-oneill.github.io/localbubble/density/}{\includegraphics[width=0.97\textwidth]{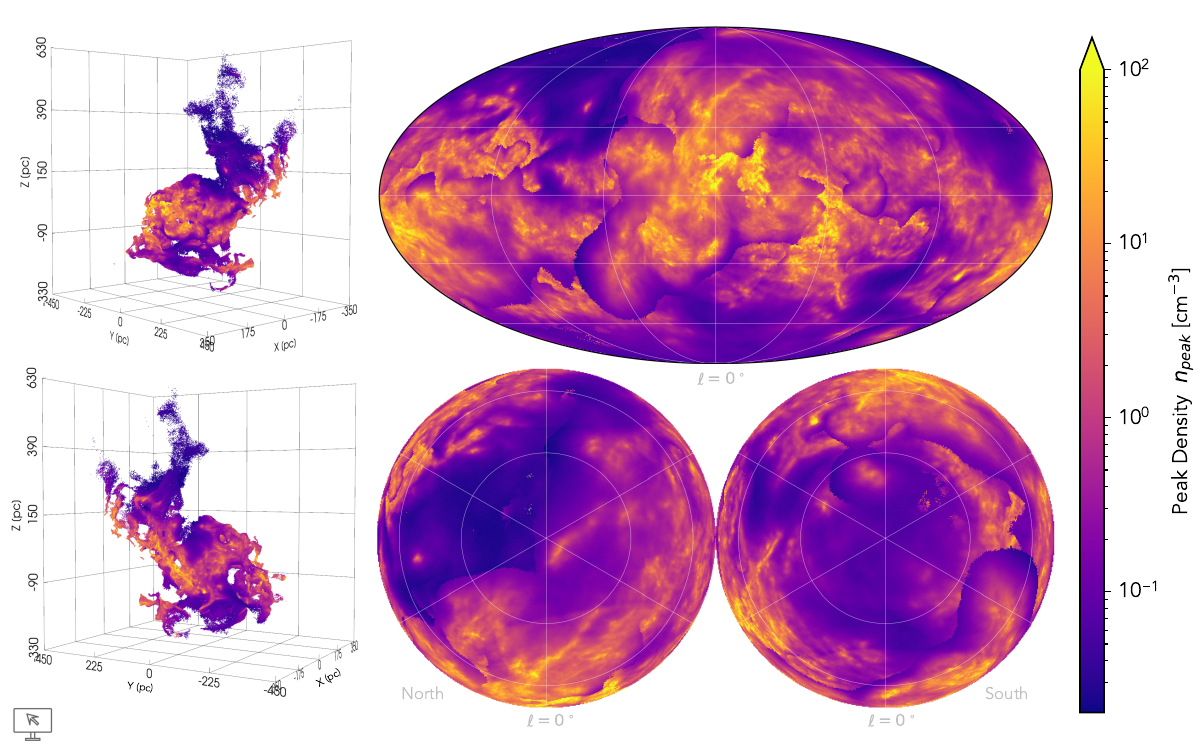}}
    \caption{Peak density of dust in the \LB's shell, $n_{peak}$, in 3D and 2D projections.  \underline{Interactive 3D figure}: \url{https://theo-oneill.github.io/localbubble/density/}}
    \label{fig:dens_peak}
\end{figure*}

\begin{figure*}
     \centering
     \href{https://theo-oneill.github.io/localbubble/thickness/}{\includegraphics[width=0.97\textwidth]{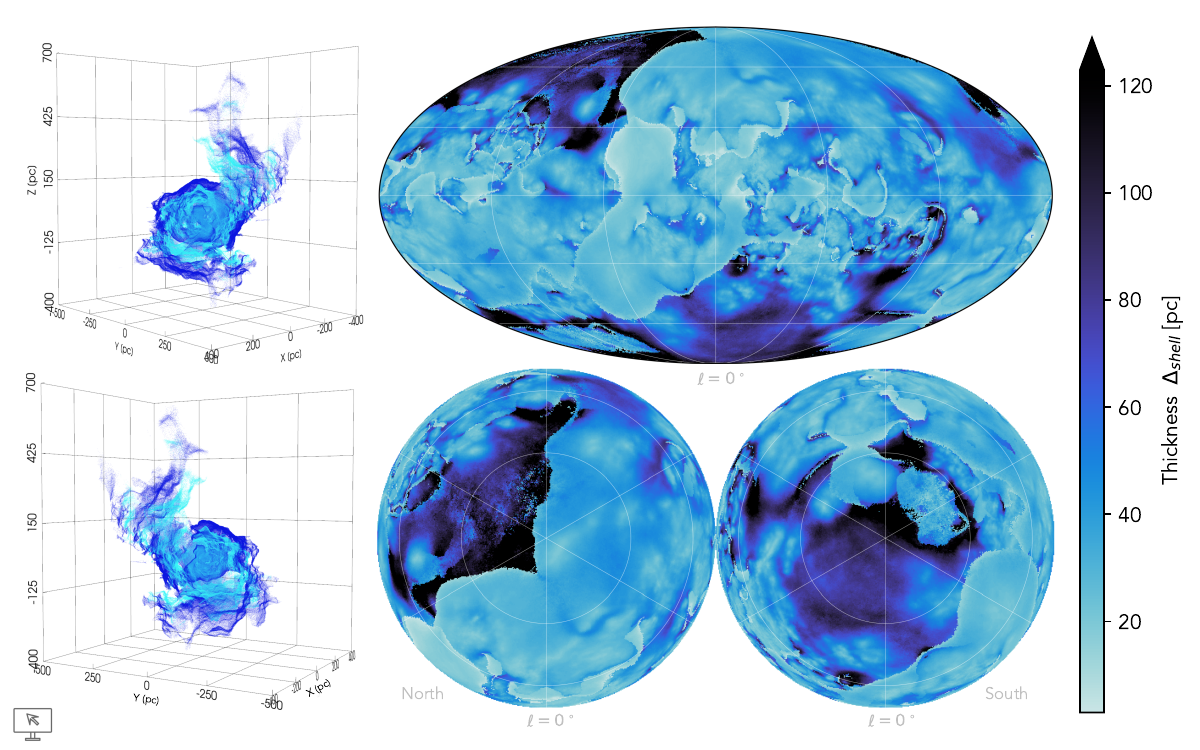}}
     \caption{Thickness of the \LB's shell, $\Delta_{shell}$, in 3D and 2D projections.  In 3D, the inner surface is shown by the light blue points and the outer surface by the dark blue points.  \underline{Interactive 3D figure}: \url{https://theo-oneill.github.io/localbubble/thickness/}}
     \label{fig:thick_peak}

    \href{https://theo-oneill.github.io/localbubble/inclination/}{\includegraphics[width=0.97\textwidth]{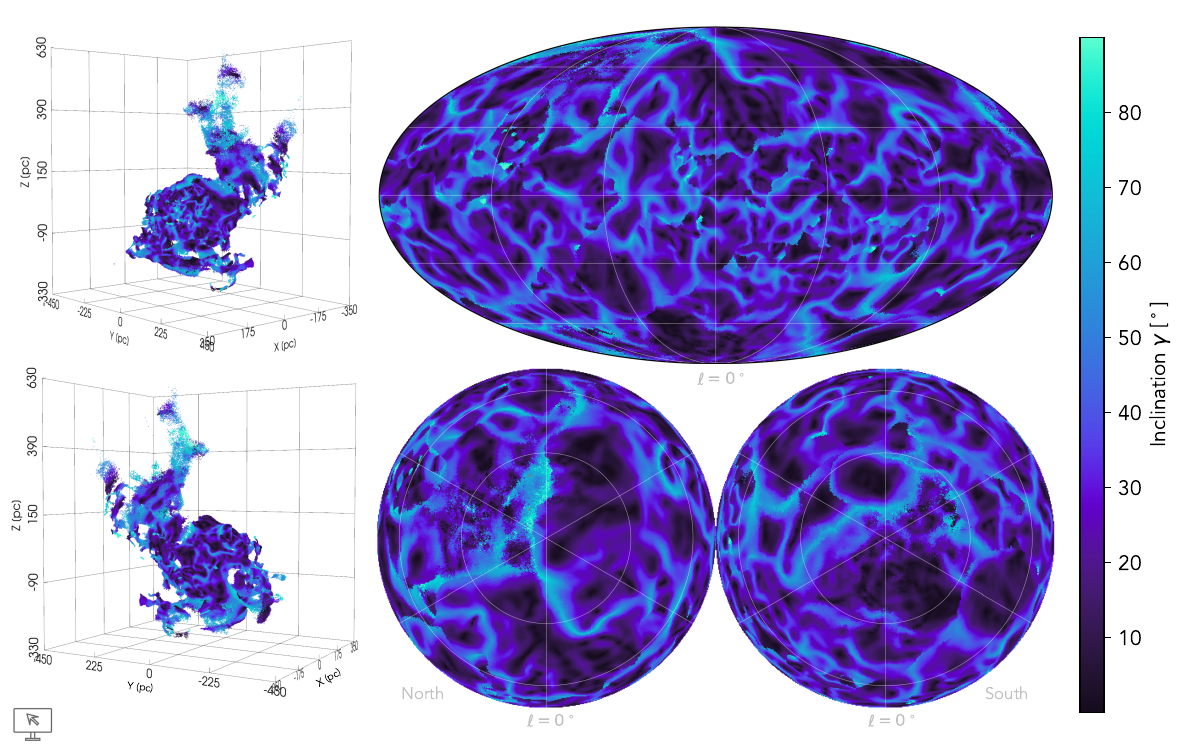}}
    \caption{Inclination of the \LB's shell to the plane-of-the-sky, $\gamma$, in 3D and 2D projections. \underline{Interactive 3D figure}: \url{https://theo-oneill.github.io/localbubble/inclination/}}
    \label{fig:gamma_peak}
\end{figure*}

\begin{figure*}
     \centering
     \includegraphics[width=\textwidth]{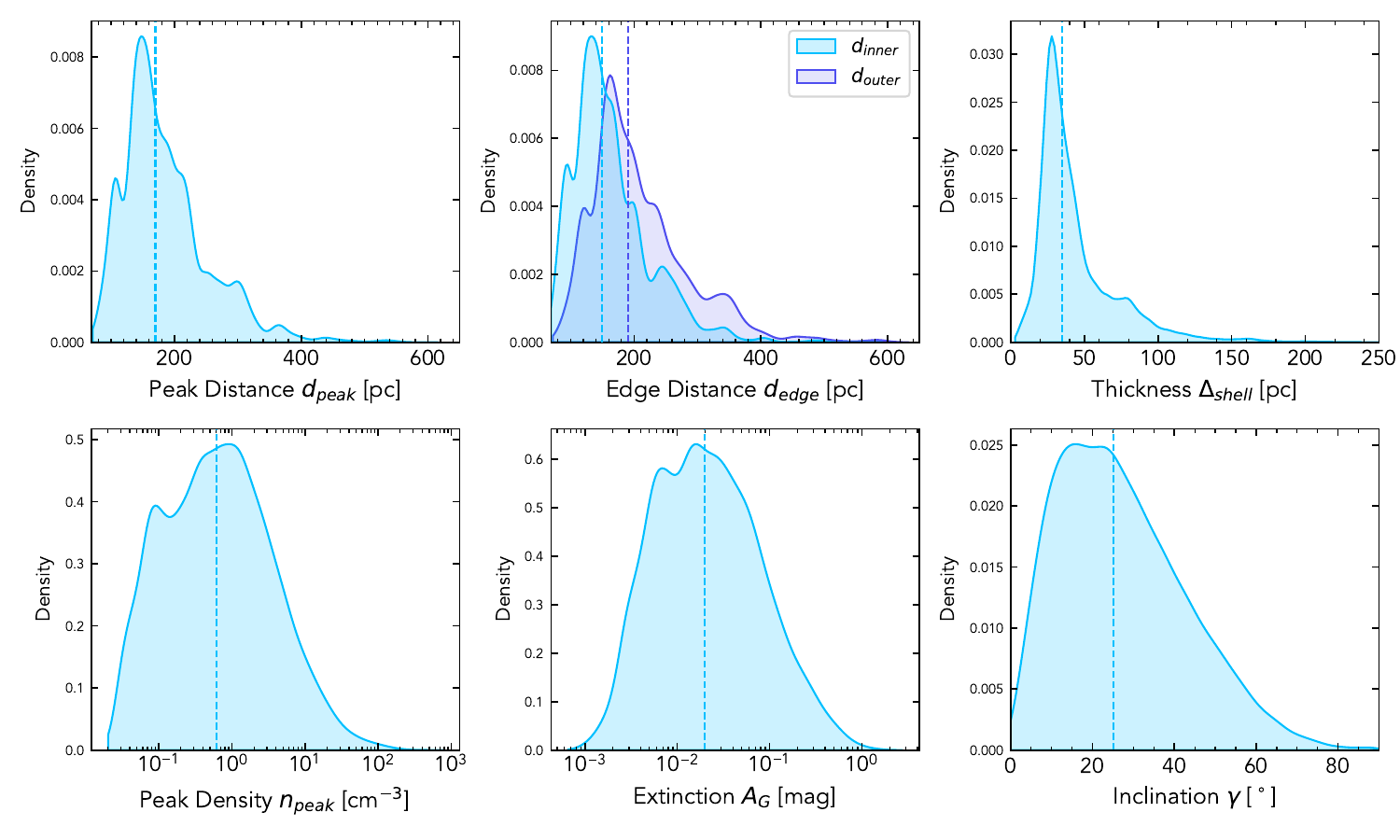}
     \caption{Kernel density estimations of \LB shell properties.  \textit{Top, from left:}  peak distance $d_{peak}$; inner and outer edge distances \dinner and \douter; thickness $\Delta_{shell}$.  \textit{Bottom, from left:} peak density $n_{peak}$; extinction $A_G$; inclination to the POS $\gamma$.  In all panels, the median of the distribution is marked by a dashed vertical line.}
     \label{fig:shell_stats}
\end{figure*}

\begin{figure}
     \centering
     \includegraphics[width=0.5\textwidth]{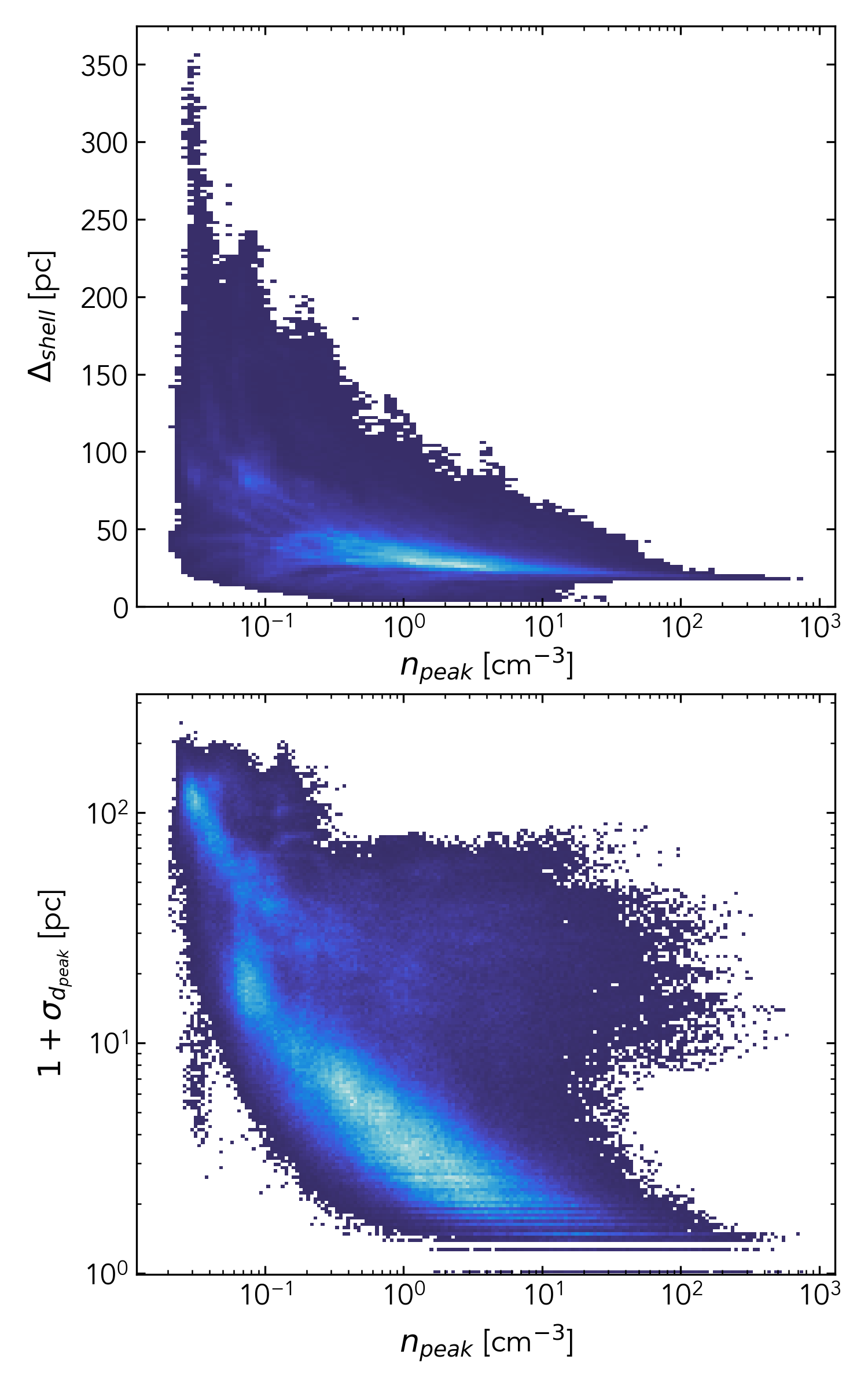}
     \caption{2D histograms of shell peak density $n_{peak}$ vs. (\textit{top}) shell thickness $\Delta_{shell}$, and vs. (\textit{bottom}) peak distance uncertainty $\sigma_{d_{peak}}$ (plotted as $1 + \sigma_{d_{peak}}$ to allow a logarithmic scale without excluding LOS with $\sigma_{d_{peak}}=0$ pc). }
     \label{fig:shell_corr}
\end{figure}

Dust identified as being a part of the \LB's shell by our peak-finding procedure is highlighted in color in Figure \ref{fig:slices_2d} in interactive 3D form and in  2D slices through the $x=0$ pc, $y=0$ pc, and $z=0$ pc planes; the rest of the E24 map is shown in grayscale.  Various properties of the \LB's shell are shown in 2D projection and 3D interactive form in the subsequent figures. Distance from the Sun to the peak extinction surface of the \LB, $d_{peak}$, in shown in Figure \ref{fig:distance_peak}, and peak shell density, $n_{peak}$, is shown in Figure \ref{fig:dens_peak}.   Shell thickness, $\Delta_{shell}$, (at $A'_{0.5}$) is shown in Figure \ref{fig:thick_peak}, and shell inclination to the POS, $\gamma$, is shown in Figure \ref{fig:gamma_peak}.  Uncertainties on peak distance, $\sigma_{d_{peak}}$ are shown in Appendix \ref{ap:uncertainty}'s Figure \ref{fig:sig_distance_peak}.  Figure \ref{fig:topdown_props} in Appendix \ref{ap:topdown} shows additional static 3D views of Figures \ref{fig:distance_peak}--\ref{fig:gamma_peak} from ``above'' and ``below.'' The statistical distributions of the shell's properties are summarized in Table \ref{tab:shell_props} and shown graphically in Figure \ref{fig:shell_stats}.

Our model of the \LB's shell is extremely irregular and asymmetric, with wide variations in morphology and peak properties present over the surface of the Bubble.  At low altitudes, the shape of the \LB is roughly spherical, but is marked by many small scale discontinuities and extensions.  At higher altitudes, a large scale, asymmetric region of increased distances and decreased densities is present towards Galactic North; we discuss this feature (which we associate with representing a Chimney out of the Galactic plane) in \S\ref{S:chimney}.  

The \LB's shell is much more extended in the $y$ and $z$ directions than in $x$, spanning the coordinates $-300 \leq x \leq 330$ pc, $-355 \leq y \leq 445$ pc, and $-300 \leq z \leq 600$ pc.  The geometric center of the \LB's shell is located at $(x, y, z) = (-9.8 \pm 0.3, 1.9 \pm 0.3, 0.2 \pm 0.8)$ pc, although we caution that our peak-finding method extending radially outward from the Sun's position at $(x, y, z) = (0, 0, 0)$ pc likely influences this result.  We additionally emphasize that our minimum distance to the shell's inner edge of $d_{inner} = 69$ pc is set by the undefined inner region of the E24 map, and note that it is possible that portions of the shell extend within this inner region.  However, previous models of the \LB generated from alternative methods and data (summarized in \S\ref{S:intro}) provide strong evidence that this interior region is mostly empty (with the exception of the ``Local Fluff'' cloud complex, which has a typical \HI{} number density of $n=$0.1--0.2 cm$^{-3}$, \citealt{LinskyRedfield2022}), and is therefore likely not relevant to our goal of mapping the \LB's dense shell.

Various correlations exist between properties of the \LB's shell.  As shown in Figure \ref{fig:shell_corr}, thickness and peak density are moderately negatively correlated (Spearman $\rho_s = -0.61$, $p < 0.001$), with thicker shell sections generally having lower peak densities.  Uncertainties on distance to the shell's surface are also moderately negatively correlated with peak density ($\rho_s = -0.66$, $p < 0.001$), with denser structures having lower distance variability between draws of the E24 map.  

We find a total dust-traced mass in the \LB's shell of $M_{shell} = (6.0 \pm 0.7 \ (\textrm{systematic}) \pm 0.1 \ (\textrm{statistical})) \times 10^5 \ M_\odot $ at an edge threshold of $A'_{0.5}$.  As described in \S\ref{S:uncert_prop}, this quantity depends strongly on the size of the smoothing kernel applied along the LOS; for $\sigma_{smooth}=6$ pc, $M_{shell} = 5.6 \times 10^5 \ M_\odot$, while for $\sigma_{smooth}=8$ pc, $M_{shell} = 6.3 \times 10^5 \ M_\odot$.  At a more generous threshold of $A'_{0.9}$, the total mass (for the fiducial $\sigma_{smooth}=7$ pc) is $M_{shell} = 8.0 \times 10^5 \ M_\odot $.  These estimates include the dust-traced mass of a number of nearby molecular clouds (discussed further in \S\ref{S:clouds}).  At all thresholds, this shell mass measurement is smaller than \citet{ZuckerGoodman2022}'s estimate of $M_{shell} = (1.4 \pm 0.6) \times 10^6 \ M_\odot$ derived from the \citet{PelgrimsFerriere2020} model of the \LB (assuming a typical shell thickness of 50--150 pc) using the \citet{LeikeGlatzle2020} 3D dust map.  We attribute this difference to the decreased thickness of the \LB in our new model, enabled by the increased spatial resolution of the E24 dust map.

\subsection{Properties of the Local Bubble's Interior}

We additionally derive several relevant characteristics of the interior of the \LB (defined for our purposes as $d < d_{inner}(A'_{0.9})$).  The \LB's interior covers a volume of $1.9 \times 10^7$ pc$^{3}$ (including the $d<69$ pc sphere that the E24 map does not probe), which is equivalent to a sphere with a radius of $R = $165 pc.  

The volume density of dust within the \LB's interior (excluding the inner $d < 69$ pc region) spans a $\mp 2\sigma$ interval from $n = 0.005$ cm$^{-3}$ to $n = 0.35$ cm$^{-3}$, with a median of $n=$0.02 cm$^{-3}$.  This median density of $n=$ 0.02 cm$^{-3}$ is roughly consistent with the predicted density derived from the pulsar dispersion measure. \citet{LinskyRedfield2021} find a typical electron density of $n_e = 0.0120 \pm 0.0029$ cm$^{-3}$ for lines of sight towards the nearest five pulsars from the Sun, lying at distances of 156-372 pc. Assuming the Local Bubble's interior is fully ionized ($n_e = n_H^{+}$), the volume density of hydrogen nuclei $n$ we derived in Eqn. \ref{eqn:vol_dens} would be equal to $n_H^{+}$, so our derived density agrees with the density predicted by the pulsar dispersion measure to within a factor of two.

Along each LOS, the ratio of the minimum dust density in the \LB's interior, $n_{min, interior}$, to the peak dust density in the \LB's shell, $n_{peak}$, spans a $\mp 2 \sigma$ interval from $n_{peak}/n_{min, interior}=$2 to 800, with a median ratio of $30$.  The ratio between the interior and shell dust density is lowest along the Northern Chimney feature and along the diffuse edges of the lower-latitude shell.

\section{Discussion}\label{S:discuss}

\begin{figure*}
    \centering
    \href{https://theo-oneill.github.io/localbubble/neighborhood/}{\includegraphics[width=\textwidth]{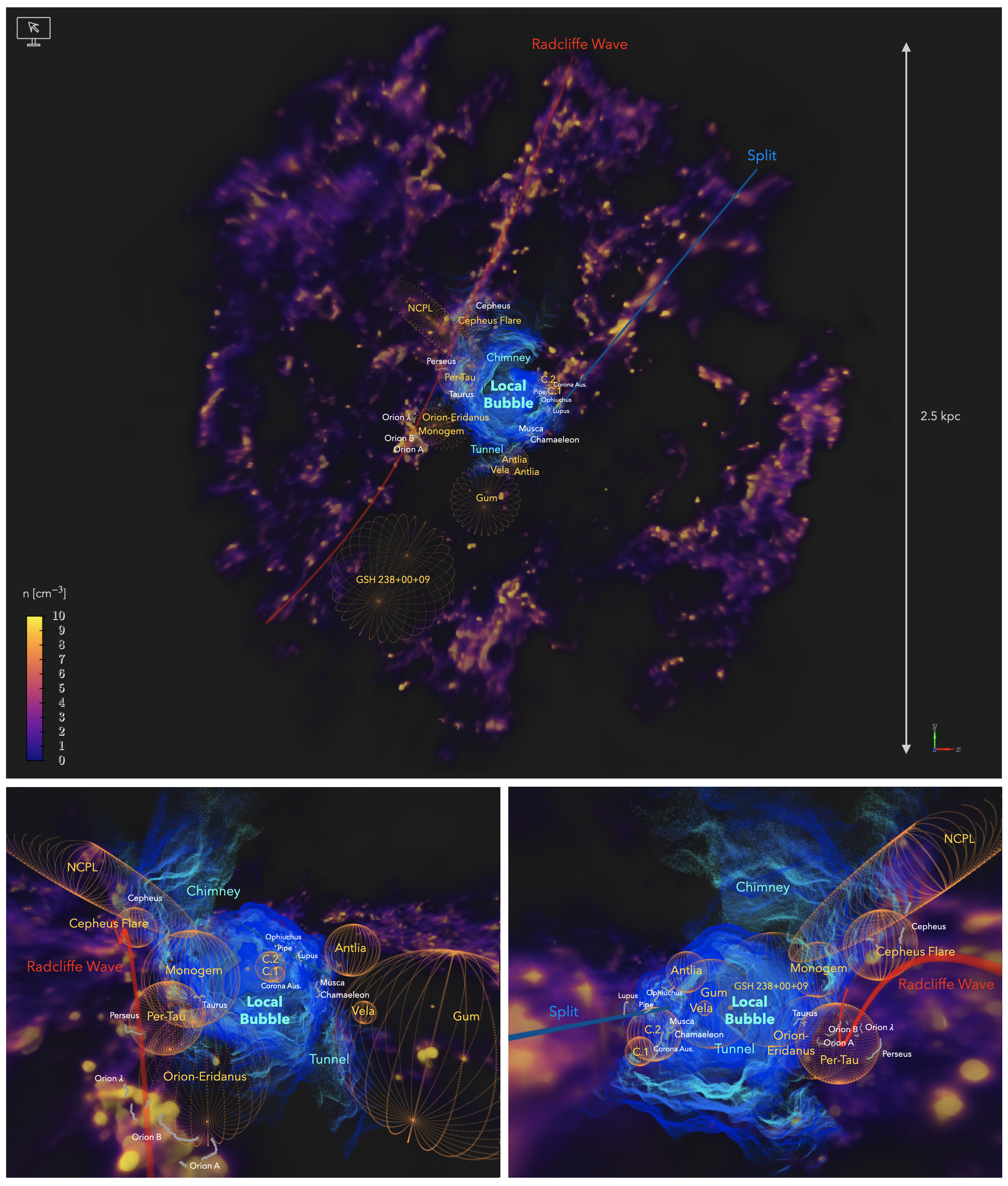}}
    \caption{A 3D interactive view of the \LB in the context of the Solar Neighborhood.  The inner, peak, and outer edges of the \LB's shell are shown in light blue, medium blue, and dark blue, respectively.  Structures in the Solar Neighborhood (including 3D dust, nearby bubble and shells, molecular clouds, and other Galactic features) are summarized in Table \ref{tab:summary_3ddat}.  \underline{Interactive 3D figure}: \url{https://theo-oneill.github.io/localbubble/neighborhood/}}
    \label{fig:enviro}
\end{figure*}

\begin{figure*}
    \centering
    \includegraphics[width=0.75\textwidth]{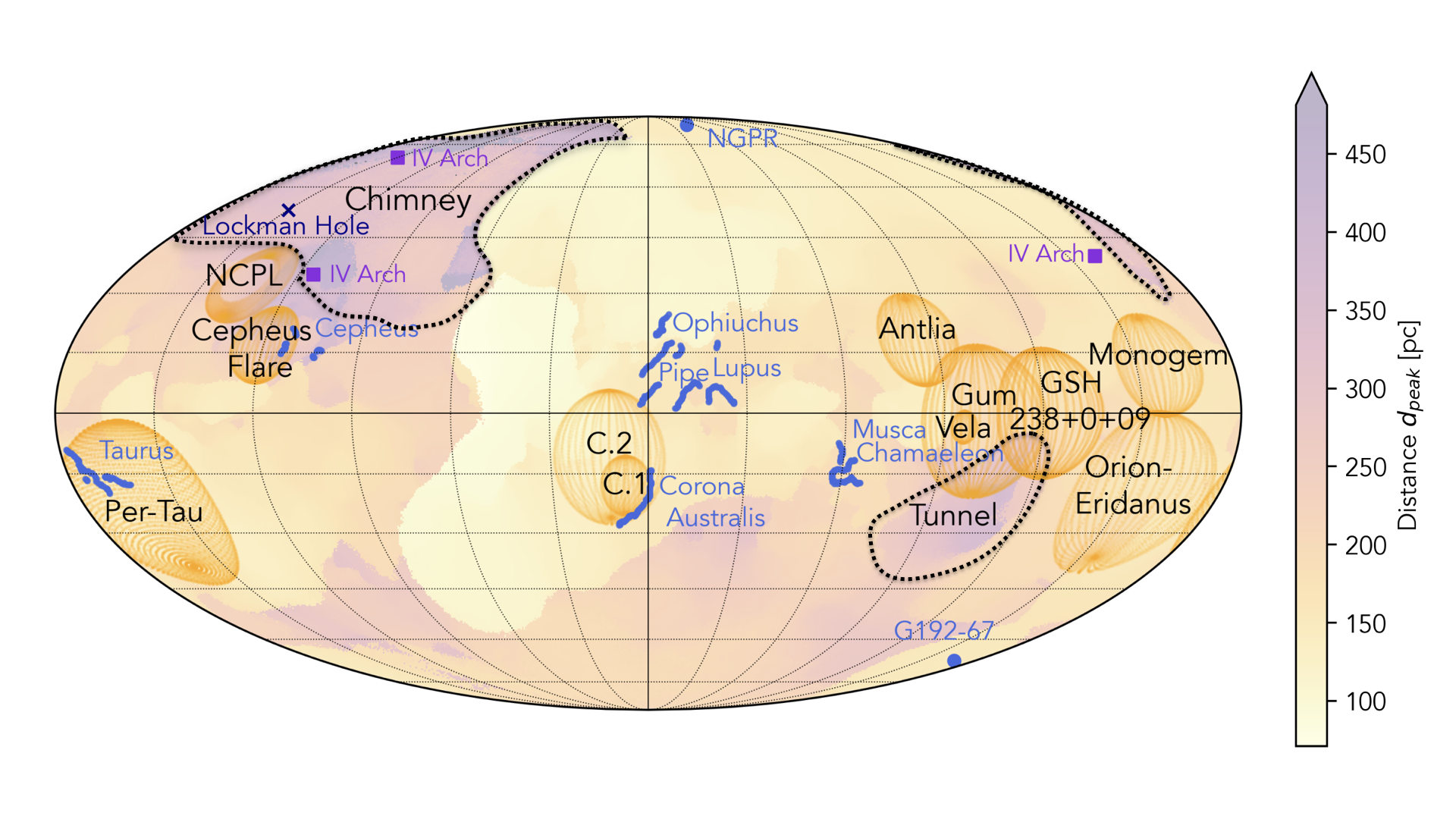}
    \caption{A 2D view of structures in the Solar Neighborhood that may be relevant to the \LB's evolution.  This includes nearby bubbles and shells (orange), molecular clouds and dust features (blue), and other features of the 2D sky discussed in text.  Distance to the \LB's shell (Fig. \ref{fig:distance_peak}) is shown in the background.  Approximate outlines of the ``Chimney'' and ``Tunnel'' features are shown by the black dotted contours, and are drawn by eye based on shell distance and density.}
    \label{fig:enviro_2d}
\end{figure*}

\begin{deluxetable*}{c|c|c|c}
\centering
\tablecaption{Structures in the Solar Neighborhood \label{tab:summary_3ddat}} 
\tablehead{\colhead{Object Type}	&	\colhead{Name}	&	\colhead{Notes} & \colhead{Reference}}
\startdata	
\textbf{Local Bubble} & Local Bubble Model & Peak Density, Inner \& Outer Edges & This work \\
\hline
\textbf{Dust} & 3D Dust Map & Downsampled to (10 pc)$^3$ voxels & \citet{EdenhoferZucker2024} \\
\hline
\multirow{2}{2cm}{\centering \textbf{Molecular Clouds}}  & \multirow{2}{2.25cm}{\centering Cloud Skeletons} & Cepheus, Chamaeleon, Corona Australis, Lupus, & \multirow{2}{2.8cm}{\centering \citet{zucker2021}} \\ 
 & & Musca, Ophiuchus, Orion, Perseus, Pipe, Taurus & \\
\hline
\multirow{2}{2cm}{\centering \textbf{Galactic Structures}}  & Radcliffe Wave & & \citet{KonietzkaGoodman2024} \\
 & Split$^{a}$ &  & \citet{LallementBabusiaux2019} \\
\hline
\multirow{11}{2cm}{\centering \textbf{Shells \& Cavities}} & Per-Tau Shell  & (x, y, z, R) = (-190, 65, -84, 78) pc  & \citet{BialyZucker2021} \\
& NCPL & Prolate Spheroid & \citet{MarchalMartin2023} \\
\cline{2-4}
& Antlia SNR & ($\ell$, $b$, $d$, D) = (275.5$\degree$, 18.4$\degree$, 250 pc, 23$\degree$) & \citet{FesenDrechsler2021} \\
& C.1 Shell$^{b}$ & ($\ell$, $b$, $d$, D) = (7$\degree$, -18$\degree$, 190 pc, 15$\degree$) & \citet{BraccoBresnahan2020} \\
& C.2 Shell$^{b}$ & ($\ell$, $b$, $d$, D) = (12$\degree$, -11$\degree$, 160 pc, 33$\degree$) & \citet{BraccoBresnahan2020} \\
& Cepheus Flare Shell & ($\ell$, $b$, $d$, D) = (120$\degree$, 17$\degree$, 300 pc, 19$\degree$) & \citet{OlanoMeschin2006} \\
& Gum Nebula & ($\ell$, $b$, $d$, D) = (258$\degree$, -2$\degree$, 400 pc, 36$\degree$) & \citet{SushchHnatyk2011} \\
& \GSHbub & ($\ell$, $b$, $d$, D) = (238$\degree$, 0$\degree$, 790 pc, 32$\degree$) & \citet{Heiles1998} \\
& Monogem Ring & ($\ell$, $b$, $d$, D) = (203$\degree$, 12$\degree$, 300 pc, 25$\degree$) & \citet{KniesSasaki2018} \\
& Orion-Eridanus & ($\ell$, $b$, $d$, D) = (205$\degree$, -20$\degree$, 290 pc, 40$\degree$) & \citet{PonOchsendorf2016} \\
& Vela SNR & ($\ell$, $b$, $d$, D) = (264$\degree$, -3.4$\degree$, 290 pc, 8$\degree$) & \citet{SushchHnatyk2011} \\
\enddata	
\tablecomments{The models of the Per-Tau Shell and the NCPL were derived from 3D dust maps.  The other shells' centers are estimated from 2D data and projected to 3D based on assumed distances $d$ and on-sky diameters $D$.  \\
$^{a}$ The linear model of the Split is placed by eye as it agrees with the E24 map at a uniform altitude of $z=0$ pc, and is intended only to guide the viewer to the relevant dust feature.  \\
$^{b}$ The distances derived for the C.1 and C.2 shells by \citet{BraccoBresnahan2020} ($d = 186$ pc and $d= 144$ pc, respectively) were measured at the edges of the shells as viewed in 2D projection; in this work, we place the shells' centers at estimated distances that are one solution geometrically consistent with the measured distances to the shell edges.}
\end{deluxetable*}

We provide a 3D interactive view of the \LB in the context of the Solar neighborhood in Figure \ref{fig:enviro}.  Our model of the \LB is represented through its inner and outer edges, as well as position of peak extinction along the LOS.  Structures represented in this figure are listed in Table \ref{tab:summary_3ddat} and discussed throughout this section; a subset of structures that may be adjacent to the \LB's surface are shown in 2D projection in Figure \ref{fig:enviro_2d}. 

We highlight several features of our new model of the \LB:

\subsection{A Local Chimney}\label{S:chimney}

The most prominent feature of our new model is that the Northern cap of the \LB appears partially open, being marked by a region of particularly distant ($d =$300--600 pc) and low-density dust.  This open cap spans about $\frac{1}{3}$ of the high latitude ($b > 30\degree$) Northern sky, extending over the range of $\ell \simeq$ ($60\degree$ -- $190\degree$) and $b \gtrsim 30\degree$.  The density of material in this feature is the lowest over the entire sky as we have estimated it (typically $\sim 10^{-1.5}$ cm$^{-3}$).  Shell thickness is also extremely high (>100 pc).  In 3D space, this region of extremely low density dust spans altitudes of $z \simeq 150$ pc to $z \simeq 600$ pc (although we emphasize that distance and, correspondingly, altitude uncertainties are generally very large in this region as a result of the low densities being probed).

Early 3D maps of the \LB \citep[e.g.,][]{SfeirLallement1999,WelshSfeir1999, VergelyFreireFerrero2001, LallementWelsh2003} proposed that the \LB was a \LC, with tilted, open caps to both the North and South.  More recent models of the \LB made with intermediate-resolution dust maps \citep{PelgrimsFerriere2020} have found that the Northern and Southern caps both appeared closed.  The distant, low-density Northern dust feature in our new model appears morphologically consistent with supporting the earlier view of a \LC extending from the \LB into the lower Galactic halo.  The tilt of the Northern Chimney in our model of the \LB (centered roughly towards \ellb{120}{60}) is similar but not identical to the tilt found in those earlier \LC maps like that of \citet{LallementWelsh2003}, who reported a tilt centered towards \ellb{180}{70}.  
 
Unlike the earlier maps, we find a closed surface across the Southern cap of the \LB.  This surface appears relatively cohesive and traces a nearly flat 3D surface comprised of dust that is generally higher density ($\simeq 10^{-1}$ cm$^{-3}$) than the tenuous Northern material (but is still lower density than the bulk of the lower latitude shell).  Uncertainties on distance to this material are much lower than in the low density Northern material, and thickness and shell inclination are generally more consistent between adjacent LOS than in the North.  This suggests that the Southern cap is closed, and that the \LB is an asymmetric \LC.

An asymmetric \LC presents interesting implications for the past and present evolution of the Solar Neighborhood.  Over the last several decades, a variety of analytic models \citep[e.g.,][]{KooMcKee1992,BaumgartnerBreitschwerdt2013,OrrFielding2022,OrrFielding2022_turbulence} and numerical treatments \citep[e.g.,][]{MacLowMcCray1988, FerriereMacLow1991,deAvillezBerry2001,KimOstriker2017,KimOstriker2018} exploring factors related to bubble breakout have emerged, including 2D hydrodynamical \citep{MacLowMcCray1989}, 3D hydrodynamical \citep{FieldingQuataert2017,FieldingQuataert2018}, and 3D magneto-hydrodynamical \citep{Tomisaka1998,KorpiBrandenburg1999,deAvillezBreitschwerdt2005,WalchGirichidis2015,GirichidisWalch2016} simulations, spanning a wide range of initial disk conditions, box sizes, and run times.  These works generally suggest that superbubbles are only able to break out of the denser gas in the disk and form chimneys and fountains reaching into the Galactic halo under specific conditions; the stratified distributions of gas density, magnetic field orientation \& strength, and related factors are found to play large roles in inhibiting or encouraging superbubble blowout, as are the positions of a bubble's progenitor supernovae relative to the midplane of the disk.

Assuming the Galactic midplane falls somewhere between $z\simeq$ -25 to -5 pc \citep[e.g.,][relative to the Sun's position and IAU midplane at $z=0$ pc]{MaizApellaniz2001_spatial,JuricIvezic2008,AndersonWenger2019}, the vertical extrema of the \LB falls $>$600 pc above the midplane in the North and $>$270 pc below the midplane in the South.  The scale height of \HI{} gas in Milky Way-like galaxies is expected to flare with galactocentric radius and be of order a few hundred parsecs near the position of the Sun \citep{BacchiniFraternali2019, Patra2020, GensiorFeldmann2023}, placing the Northern extremes of the \LC in the lower halo.  We caution that the high uncertainty, low density dust found in this region should not be interpreted as a definitive detection of an upper boundary or ``end'' of the Northern \LC.  

The observed asymmetric \LC suggests that either the physical conditions in the Galactic Northern and Southern hemispheres were significantly different from each other before the expansion of the \LB, or that the \LB's progenitor (or even only most recent) SN were preferentially located in the North, above the Galactic midplane.  Blow out in the South might then have been halted by encountering the denser gas in the plane, while blow out in the North would have been able to proceed due to a lack of interference.  The Sco-Cen association is a strong candidate for hosting the progenitor clustered supernovae that launched the initial expansion of the \LB \citep[as proposed by, e.g.,][]{MaizApellaniz2001,FuchsBreitschwerdt2006,BreitschwerdtFeige2016}; stellar tracebacks calculated by \citet{ZuckerGoodman2022} place clusters in the association at heights of only $z =$ -16 to -17 pc when the Bubble was born $\sim$14 Myr ago, roughly consistent with the height of the present-day midplane.  

In their analytic modeling of the effects of clustered supernovae on bubble evolution, \citet{OrrFielding2022_turbulence} draw a distinction between superbubbles that simply coast out of the disk (meaning that the production of the supernovae powering a bubble's expansion has ceased by the time the bubble reaches the disk scale height) vs. those whose breakout is powered by ongoing supernovae.  They find that coasting superbubble breakout should be extremely difficult to achieve under typical conditions of disk galaxies like the Milky Way.  In this context, the apparent bursting morphology of the \LB suggests that the \LC was formed from the \LB under the power of ongoing supernovae -- although we note the time and position of the most recent supernovae in the \LB is uncertain, and the interval elapsed since the \LB formed a chimney is unknown.  

A \LC extending from the \LB would additionally have implications for the present-day conditions in the interior of the \LB, especially temperature and pressure, as chimneys are expected to play a key role in dissipating the hot interiors of feedback-driven bubbles.  Initial evidence for a hot \LB has faced challenges in recent years \citep[e.g.,][]{WelshShelton2009,LinskyRedfield2021}, and a chimney venting the initially hot Bubble interior into the lower halo could provide a path to a resolution of this long standing issue.  However, the degree to which mixing and cooling could have occurred would directly depend on the timescale of the \LB's breakout and formation of a chimney, which at present is unconstrained.

\subsubsection{Connection to Intermediate Velocity Clouds?}

In addition to allowing the hot interior of superbubbles to be dissipated, chimneys from burst superbubbles are theorized to lead to the creation of populations of intermediate velocity clouds (IVCs) contributing to a galactic fountain flow.  In this model, IVCs are posited to be fragments of burst superbubbles that were launched into the lower Galactic halo, before cooling and raining back down onto the Galactic plane --- propagating a galactic fountain and leading to new star formation and the distribution of metals throughout the ISM \citep{ShapiroField1976,Bregman1980}.

If the \LB is only open to the North as found in our new model, one might expect to observe a higher number of IVCs in the North than the South, with distinctions arising both chemically and kinematically between the populations of IVCs.  Such an asymmetry between the Northern and Southern distributions of IV gas is well known, with a number of studies having found significant differences in the distribution of IVCs between hemispheres.  Much of this asymmetry is driven by large Northern complexes of IV gas with negative radial velocities known as the IV Arch, IV Spur, and low latitude IV Arch \citep{KuntzDanly1996}.  These features are easily observable in observations of IV 21 cm emission in the Northern vs. Southern hemispheres \citep[e.g., as compiled by][]{AlbertDanly2004,Wakker2004}; the IV Arch in particular stretches from approximately \ellb{115}{35} to \ellb{150}{70} to \ellb{200}{40} \citep{KuntzDanly1996}, as shown by reference points in Figure \ref{fig:enviro_2d}. 

Statistical analyses of the Northern vs. Southern IVC populations have confirmed the asymmetries visible by eye.  \citet{RohserKerp2016} found a strong asymmetry in the number of high-latitude ($|b| \geq 20\degree$) IVCs with molecular gas content in the Northern vs. Southern hemispheres, with the North having 3.6$\times$ the number of molecular IVCs detected in the South.  \citet{PanopoulouLenz2020} similarly observed a very strong asymmetry in the total number of IVCs along each LOS in the North vs. South (on average, 3 clouds per \Hpx pixel in the North vs 2.5 clouds per pixel in the South), as assessed via Gaussian decomposition of high latitude ($|b| \geq 30\degree$) \HI{} line data compiled by the HI4PI survey \citep{HI4PICollaborationBenBekhti2016}.  Many of these observed statistical asymmetries are in part driven by the presence of the IV Arch and adjacent IV gas. 

We observe a correspondence between the on-sky position of the IV Arch (and adjacent IV complexes) and the projected low-density \LC, and speculate that there could be a connection between the two features.  A relationship between the (symmetric) \LC and general IVC population was briefly proposed by \citet{LallementWelsh2003}, and specifically between the \LC and the IV Arch by \citet{WelshSallmen2004} (and, even before the \LB was referred to as a Chimney, a connection between the \LB and the Northern infalling IV \HI{} complex was suggested by \citealt{DickeyLockman1990}).  However, we note that the \LB is not the only superbubble in the Solar Neighborhood (see \S\ref{S:tunnel}) and is just one potential contributor among many to the larger-scale Galactic fountain, making disentangling which mechanism launched any particular IVC a difficult task.  

In addition, the specific mechanisms that may influence the creation of IVCs from burst superbubble interiors remain an active area of research, with factors such as the timescale for vented hot gas to cool within the halo, the effects of magnetic fields, and the overall efficiency of IVC formation \citep[e.g.,][]{Kahn1991,KahnBrett1993,Wakker2001,Fraternali2017,LehnerHowk2022,MarascoFraternali2022} all likely playing important roles in determining the fraction of chimneys that launch IVCs.  Future work should investigate the potential relationship between the \LC and the broader IVC population via detailed modeling of the many physical processes influencing both superbubble blowout and IVC formation.

\subsection{Tunnels to Other Bubbles and Voids}\label{S:tunnel}

The Local Bubble is our Solar Neighborhood's connection to the theoretical context of a multiphase ISM shaped by supernova feedback \citep{CoxSmith1974,McKeeOstriker1977}.  One conclusion of this theoretical viewpoint is that feedback-driven bubbles should be ubiquitous throughout the ISM.  

Many nearby candidate supernova remnants (SNRs), shells, and cavities that may represent local components of this theorized bubbly ISM have been identified in on-sky data.  We represent idealized spherical versions of a non-exhaustive sample of these candidates in Figure \ref{fig:enviro} by projecting the 2D structures to 3D space using assumed distances and on-sky diameters (summarized in Table \ref{tab:summary_3ddat}).  This sample of nearby shells includes the Gum Nebula and its embedded Vela SNR \citep{Gum1952, BrandtStecher1971, SushchHnatyk2011}, the Antila SNR \citep{McCulloughFields2002, FesenDrechsler2021}, the Monogem Ring \citep{PlucinskySnowden1996, KniesSasaki2018}, 
the Cepheus Flare Shell \citep{GrenierLebrun1989, OlanoMeschin2006}, and the \HI{} shells C.1 \& C.2 near Corona Australis \citep{BraccoBresnahan2020}.  Larger scale superbubbles observed in 2D with distance estimates placing them near the \LB include the Orion-Eridanus superbubble \citep{ReynoldsOgden1979,PonOchsendorf2016,SolerBracco2018,JoubaudGrenier2019} and the radio supershell \GSHbub \citep{Heiles1998}.  Distance estimates for many of these bubbles are consistent with locations on or near the surface of the \LB, although we emphasize that uncertainties on the distances and 3D morphologies of these 2D bubbles are very large (and, as in the case of e.g., Orion-Eridanus and \GSHbub, may significantly elongated and non-spherical) and that this sample is included only for visualization purposes. 

Very few candidate bubbles and shells have been identified and/or mapped in 3D (e.g., with 3D dust maps).  One of the few that has is the Per-Tau Shell \citep{BialyZucker2021,DoiHasegawa2021} between the Perseus and Taurus molecular clouds (represented by \citealt{BialyZucker2021} as an idealized sphere).  \citet{PelgrimsFerriere2020}'s model of the \LB made it apparent that the Per-Tau Shell is directly adjacent to the \LB, which remains true in our new model, and it has been theorized that the compression resulting from the intersection of these two bubbles led to star formation in Taurus \citep{ZuckerGoodman2022,SolerZucker2023}.

Another consequence of \citet{CoxSmith1974}'s theoretical perspective is the expectation that a significant fraction of the volume of the bubbly ISM should be riddled with a network of tunnels stretching between bubbles.  Early maps of the \LB identified an extension of the \LB in the direction of the star $\beta$ CMa \citep{FrischYork1983}, centered approximately at \ellb{235}{-15} with an on-sky diameter of $15\degree \times 10\degree$ \citep{Welsh1991}.  This extension was proposed to be a tunnel towards the nearby Gum Nebula \citep[e.g.,][]{Welsh1991} and/or \GSHbub \citep{LallementWelsh2003}.  

Our new model of the \LB also displays an extension in this direction, but centered closer to \ellb{250}{-20}.  The more distant side of this extension falls at a distance of $d = 300$--375 pc and opens into a void-like region of the E24 dust map that appears to correspond to \GSHbub. This extension of the \LB also overlaps in 2D projection with the lower half of the Gum Nebula, which is estimated to be located at a distance of $\sim$400 pc (\citealt{BrandtStecher1971, SushchHnatyk2011}, Gao et al. in preparation).  We speculate that this extension may represent an intersection between the \LB and the Gum Nebula similar to the intersection between the \LB and the Per-Tau Shell, and/or function as a tunnel to the larger void of \GSHbub.  The exact causal relationship between the many candidate bubbles and voids in this quadrant (the \LB, \GSHbub, the Gum Nebula, the Orion-Eridanus superbubble, the Monogem Ring, the Per-Tau Shell, the Vela SNR, and the Antlia SNR) is still unknown.

We additionally note a large discontinuity between the \LB's surface in the opposite direction across the Galactic plane, with a boundary near $\ell = 60\degree$ in 2D and near $x=+100$ pc, $y=+100$ pc in the X-Y plane.  In this region, dust identified as part of the \LB's shell in the region of $\ell > 60 \degree$ appears to continue in a shell-like structure traced by the second peak along the LOS (and so is not marked as part of the \LB's shell through our peak finding method).  This extended region curves towards Corona Australis, and the suspected location of the interacting \HI{} shells C.1 and C.2 \citep{BraccoBresnahan2020}, which is suggestive of this region representing a tunnel or similar extension.

Finally, we comment that the \LB is nested between the Radcliffe Wave \citep{AlvesZucker2020} and Split \citep{LallementBabusiaux2019}, both of which are kiloparsec-long linear features that contain a significant amount of the dust in the Solar Neighborhood.  The Radcliffe Wave may be the gas reservoir of the Local Arm of our galaxy \citep{SwiggumAlves2022}, while the Split is of unknown origin.  Much of the high density dust present in the Bubble's shell is bounded by or possibly part of these larger structures.  The identified extensions and tunnels stemming from the \LB, and the asymmetric Northern chimney, are located in the gap between the Radcliffe Wave and the Split.  It is easy to imagine a scenario where the expansion of the \LB was constrained in some directions by the dense gas and dust that formed the Radcliffe Wave and Split, while a Chimney and tunnel network were more easily able to form in the lower-density regions between these features.  

\subsection{Associated Molecular Clouds and Dust Features}\label{S:clouds}

\subsubsection{Star-forming Molecular Clouds}\label{S:molclouds}

\citet{ZuckerGoodman2022} used \citet{PelgrimsFerriere2020}'s model of the \LB to demonstrate that many nearby star-forming clouds are draped over the surface of the \LB, and that their recent star formation was likely triggered by the Bubble's expansion.  We show the positions of these clouds in Figure \ref{fig:enviro}, and, in agreement with \citet{ZuckerGoodman2022}, find that the clouds Taurus, Chamaeleon, Corona Australis, Musca, Lupus, Ophiuchus, and Pipe fall on the \LB's surface.\footnote{3D maps of these clouds created by \citet{zucker2021} using \citet{LeikeGlatzle2020}'s 3D dust map were used in \citet{ZuckerGoodman2022}'s analysis.  The \citet{LeikeGlatzle2020} map was used as a prior in building the E24 dust map, and so the nearby clouds recovered by \citet{zucker2021} from the \citet{LeikeGlatzle2020} map are generally morphologically similar to those observed in the E24 map.}  As detailed in Appendix \ref{ap:pelgrims}, our model of the \LB is generally similar to \citet{PelgrimsFerriere2020}'s model in low latitude, high density regions like these clouds, so this similarity is expected.  The extended Chimney feature in our new model additionally causes the Cepheus molecular cloud to border the surface of the \LB.  The presence of these star-forming clouds on the clumpy surface of the \LB suggests that the distribution of dense gas and young stars is self-organized by supernova feedback, as predicted by theory and simulations \citep{CoxSmith1974,McKeeOstriker1977,whitworth_fragmentation_1994,Dawson2013,InutsukaInoue2015,KimOstriker2017}.

\subsubsection{Other Dust Features}\label{S:dustfeatures}

Also visible on the surface of the \LB are a number of well-known dust features (visible in Figure \ref{fig:dens_peak}). One of the most well-studied is the North Celestial Pole Loop (NCPL), centered near \ellb{136}{32}.  \citet{MarchalMartin2023} proposed that the NCPL may be a cavity protruding from the \LB's surface and filled with warm-to-hot gas, based on evidence from \HI{} data and the \citet{LeikeGlatzle2020} 3D dust map.  A dust feature in our model centered at \ellb{135}{31} at a distance of 270--320 pc is reminiscent of the NCPL, although not an exact match to the NCPL's 2D shape; the second peak along the LOS (which appears to contribute the bulk of the Spider and Ursa Major cloud complexes, located beyond the \LB's shell) overlaps with these features and is closer to the traditional 2D projected view of the NCPL.  We hypothesize that the feature we are detecting is the nearer, lower-density side of the NCPL, with the bulk of the structure visible in integrated views located on the more distant side.  The NCPL is located at the lower edge of our Chimney feature, and so, in agreement with \citet{MarchalMartin2023}, we speculate that the formation of the NCPL may be related to the Local Bubble's expansion and/or the formation of the \LC. 

In the higher-latitude Northern hemisphere, one of the most notable features is an elongated dust filament centered at \ellb{310}{84}, referred to in the literature as Markkanen’s cloud \citep{Markkanen1979} and the North Galactic Pole Rift (NGPR). \citet{SnowdenKoutroumpa2015} previously drew a connection between the Local Bubble and the NGPR, which our model confirms.  Its peak extinction distance in the E24 map is $d=145$ pc, which is slightly farther than its previously estimated distance of $d = 98 \pm 6$ pc \citep{PuspitariniLallement2012}.  The NGPR hugs the shell of the \LB and traces the edge of the boundary between the dense Northern shell and the Chimney feature.  In the Southern hemisphere, the G192-67 molecular cloud centered at \ellb{192}{-67} is a prominent feature on the \LB's surface; \citet{LallementWelsh2003} previously commented on this cloud as being embedded within the \LB.  It lies at a distance of $d = 150$ pc in the E24 map.

In addition to notable higher-density dust features, we speculate that many of the on-sky locations where integrated gas density and dust extinction are observed to be particularly low may be related to the density of the \LB's shell.  Just above the NCPL is the well-known Lockman Hole \citep{LockmanJahoda1986}, centered at roughly \ellb{150}{53} and characterized by having the lowest known \HI{} column density in the sky.  The Lockman Hole overlaps in projection with the low-density \LC, and we additionally note a generally excellent correspondence between the location of low \HI{} sightlines across the full sky \citep[as visible in e.g., the HI4PI survey, ][]{HI4PICollaborationBenBekhti2016}, in both the Northern Chimney region and in the South across the polar cap and towards the proposed tunnel to the Gum Nebula/\GSHbub.  This suggests that the absence of dense \LB shell in these directions allowed certain regions of the sky to be particularly unobscured by foreground material.

\section{Conclusions}\label{S:conclude}

In this work, we have derived a new model of the \LB using the \citet{EdenhoferZucker2024} 3D dust map of the Solar Neighborhood.  The most notable feature of the new model is that the \LB's Northern cap appears to have burst, i.e., the \LB is a \LC.  Many well-known molecular clouds and dust features fall on the \LB's surface, and a number of tunnels to adjacent cavities in the ISM are observed.  

Distance to the \LB's shell ranges from 70 pc to 600$+$ pc, with a typical volume density in the Bubble's interior of $n  \sim 10^{-2}$ cm$^{-3}$ and in the Bubble's shell of $n \sim 1$--$10^2$ cm$^{-3}$.  We estimate the total mass of the \LB's dust-traced shell as $(6.0 \pm 0.7 \ (\textrm{systematic}) \pm 0.1 \ (\textrm{statistical})) \times 10^5 \ M_\odot$.  The Northern Chimney extending from the \LB spans vertical heights ranging from $z \sim 150$ pc to $z \sim 600$ pc, reaching into the expected boundary of the Milky Way's lower halo.  

This new model of the \LB can be used for applications ranging from investigating the history of star formation in the Solar Neighborhood, to the morphology and evolution of the local bubbly ISM, to the relationship between the Solar Neighborhood and the lower Galactic halo.  Future work will include modeling the \LB's 3D magnetic field structure (O'Neill et al., in preparation) and the structure of neighboring bubbles in the E24 dust map (O'Neill et al., in preparation).

\begin{acknowledgements}
We thank Andrew Saydjari, Doug Finkbeiner, Eric Koch, Cameren Swiggum, Annie Gao, Andrea Bracco, and Juan Soler for insightful discussions.  We thank Cameren Swiggum for guidance in using the K3d-jupyter visualization package.  
T.J.O., C.Z., and A.A.G. acknowledge support by NASA ADAP grant 80NSSC21K0634 ``Knitting Together the Milky Way: An Integrated Model of the Galaxy’s Stars, Gas, and Dust.''  T.J.O. acknowledges that this material is based upon work supported by the National Science Foundation Graduate Research Fellowship under Grant No. DGE 2140743.  G.E. acknowledges support from the German Academic Scholarship Foundation in the form of a PhD scholarship ("Promotionsstipendium der Studienstiftung des Deutschen Volkes").
\end{acknowledgements}

\vspace{2em}
{\large \textit{Interactive Figures:}} Interactive figures presented in this work can be accessed at: \url{https://theo-oneill.github.io/localbubble/}.  

\vspace{2em}
{\large \textit{Data Access:}} The derived models of the 3D Local Bubble can be downloaded from the Harvard Dataverse: \url{https://doi.org/10.7910/DVN/INB1RB}, including:
\begin{itemize} 
    \item Table of \LB shell properties at the fiducial edge threshold of $A'_{0.5}$
    \item Shell differential extinction (at $A'_{0.5}$) interpolated to a heliocentric Cartesian grid
    \item Supplementary table of shell properties at an edge threshold of $A'_{0.9}$
    \item Twelve tables of shell properties for each draw of the E24 map
    \item Table of mean shell properties derived from the twelve draws, with uncertainty estimates
    \item Standalone HTML files of all interactive figures
\end{itemize}

\software{Astropy \citep{astropy_2013,astropy_2018,AstropyCollaborationPriceWhelan2022};  Cmasher \citep{cmasher2020}; Dustmaps  \citep{M_Green_2018}; glue \citep{BeaumontGoodman2015, RobitailleBeaumont2019}; Healpy \citep{Zonca2019_healpy}; K3d-jupyter \citep{k3d_jupyter}; Matplotlib \citep{matplotlib_Hunter2007}; Numpy \citep{harris2020_numpy}; Pandas \citep{pandas_mckinney-proc-scipy-2010}; PyVista \citep{sullivan2019pyvista}; statsmodels \citep{statsmodels_seabold2010}}

\newpage
\appendix
\restartappendixnumbering

\section{Peak Finding Parameter Selection}\label{ap:peakfind_params}

\begin{figure*}
    \centering
    \includegraphics[width=\textwidth]{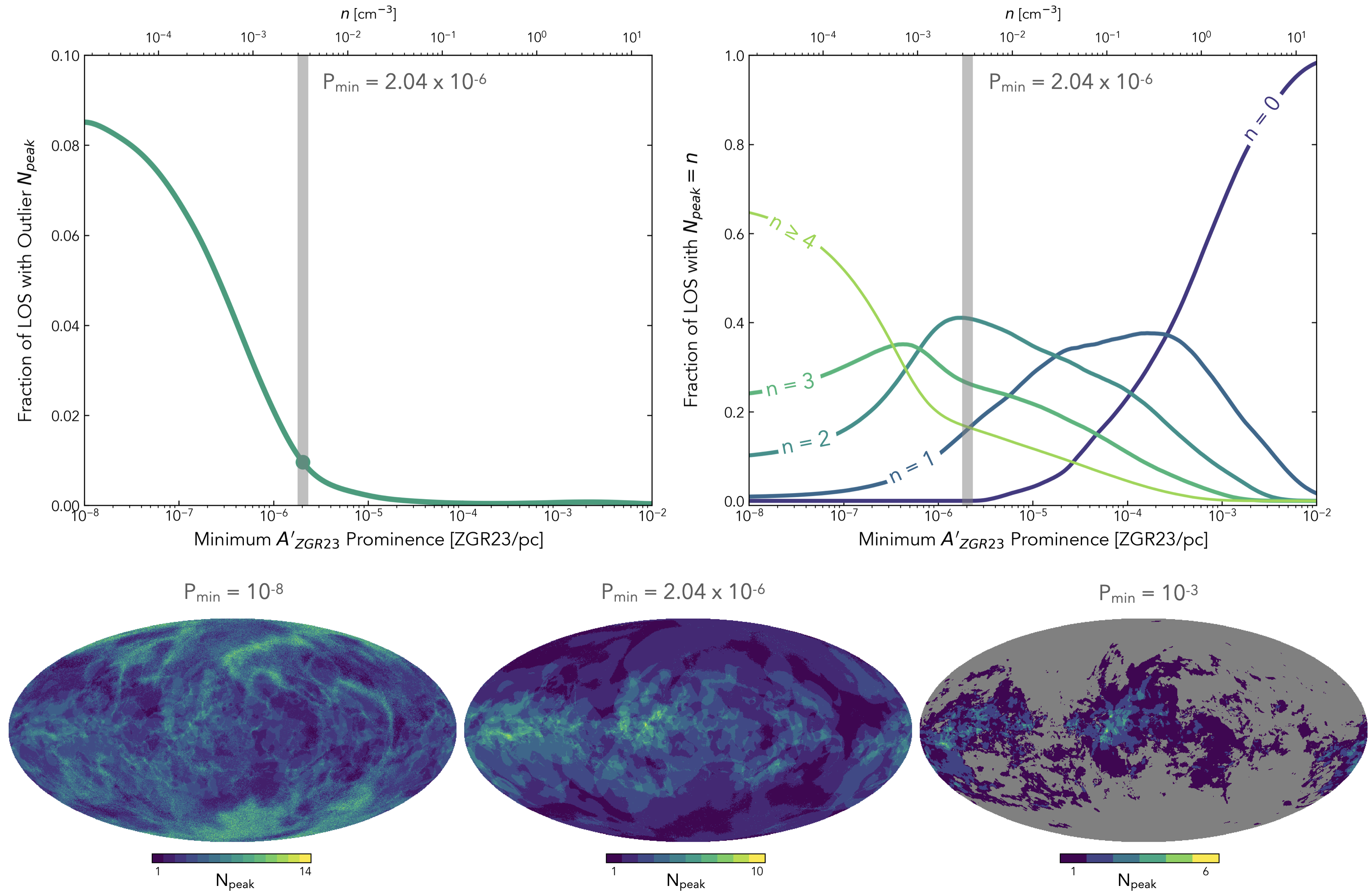}
    \caption{\textit{Top left}: Fraction of LOS with outlier numbers of peaks, \Npeak, as a function of minimum prominence \Pmin required for a local maximum along the LOS to be considered a peak.  The selected fiducial \Pmin$=2.04 \times 10^{-6}$ ZGR23/pc used to generate our model of the \LB is highlighted by the gray line.  The secondary x-axis shows the approximate conversion from ZGR23/pc to volume density $n$.  \textit{Top right:} Fraction of LOS with \Npeak$=n$ as a function of \Pmin, for $n=0, 1, 2, 3$ and $n\geq 4$.  \textit{Bottom:} Mollweide projections of \Npeak are shown for \Pmin = $10^{-8}$ ZGR23/pc, $2.0 \times 10^{-6}$ ZGR23/pc, and $10^{-3}$ ZGR23/pc.  }
    \label{fig:select_min_prom}
\end{figure*}

There are two main parameters of interest in our peak finding method: smoothing scale and minimum peak prominence.  We select smoothing scale by testing Gaussian smoothing kernel sizes ranging between $\sigma_{smooth} = 1$ pc and $\sigma_{smooth} \geq 20$ pc.  We find that $\sigma_{smooth} \lesssim 3$ pc is ineffective at smoothing out small stochastic variations in $A'_{ZGR23}$, while $\sigma_{smooth} \gtrsim 10$ pc causes narrow but distinct nearby peaks to blend together.  We find that $\sigma_{smooth} = 7$ pc generally strikes the best balance between smoothing over apparent noise while preserving genuine peaks.

We then estimate what minimum peak prominence is most suited to ignoring noise while preserving detections of high latitude, low-density peaks.  We test \Pmin ranging from $10^{-8}$ ZGR23/pc to $10^{-2}$ ZGR23/pc (with 100 values of \Pmin per dex, evenly spaced in $\log P$), spanning the broad range of differential extinction probed by the E24 map.  For each \Pmin, we performed our peak finding procedure (described in \S\ref{S:peak_find}) and calculated the number of peaks detected along each LOS, \Npeak, within a distance of $d \leq 650$ pc. 

We expect that at our well-resolved angular spacing of \nside{256}, adjacent LOS should typically have the same \Npeak, and that LOS with significantly different \Npeak from their neighboring LOS are likely to have detected some number of small-angular-scale noise-driven peaks.  Following this reasoning, we define LOS with ``outlier'' \Npeak as LOS where none of their 8 neighboring \Hpx pixels have the same \Npeak. 

For each candidate \Pmin, we perform this outlier calculation and derive the fraction of LOS over the whole sky that have outlier \Npeak.  The left panel of Figure \ref{fig:select_min_prom} shows the fraction of outlier \Npeak LOS as a function of \Pmin.  We observe a higher fraction (8\%) of outlier LOS at very low \Pmin, which gradually declines as \Pmin increases before plateauing to a low fraction ($<0.1\%$) of outlier LOS at very high \Pmin.  The bottom three panels of Figure \ref{fig:select_min_prom} show \Npeak over the sky for three representative values of \Pmin; as \Pmin increases, the on-sky distribution of \Npeak transitions from being noise-dominated, to relatively cohesive across neighboring pixels, to only detecting the highest density dust features.

The ``optimal'' \Pmin for our use case of simultaneously identifying very high density and low density peaks likely falls somewhere in the middle of this range – and, in particular, in the region experiencing the most rapid decline, ranging between \Pmin = $10^{-6}$ ZGR23/pc and \Pmin = $10^{-5}$ ZGR23/pc.  We identify this optimal \Pmin as the ``knee'' of this section of the outlier curve, and approximate the knee's location as the maximum of the second derivative of the outlier curve (smoothed with a Gaussian kernel of $\sigma = 0.1$ dex to reduce the effects of our discrete sampling of \Pmin).  This corresponds to a selected \Pmin of $2.04 \times 10^{-6}$ ZGR23/pc, a level at which only 0.9\% of LOS have outlier \Npeak.  The right panel of Figure \ref{fig:select_min_prom} shows the fraction of LOS with a given \Npeak $= n$ as a function of \Pmin for $n = 0, 1, 2, 3$ and $n \geq 4$.  At the \Pmin we selected, the majority of LOS have two or three peaks.

\section{Uncertainty Estimation}\label{ap:uncertainty}
\restartappendixnumbering

\begin{figure*}
    \centering

    \href{https://theo-oneill.github.io/localbubble/draws/}{\includegraphics[width=\textwidth]{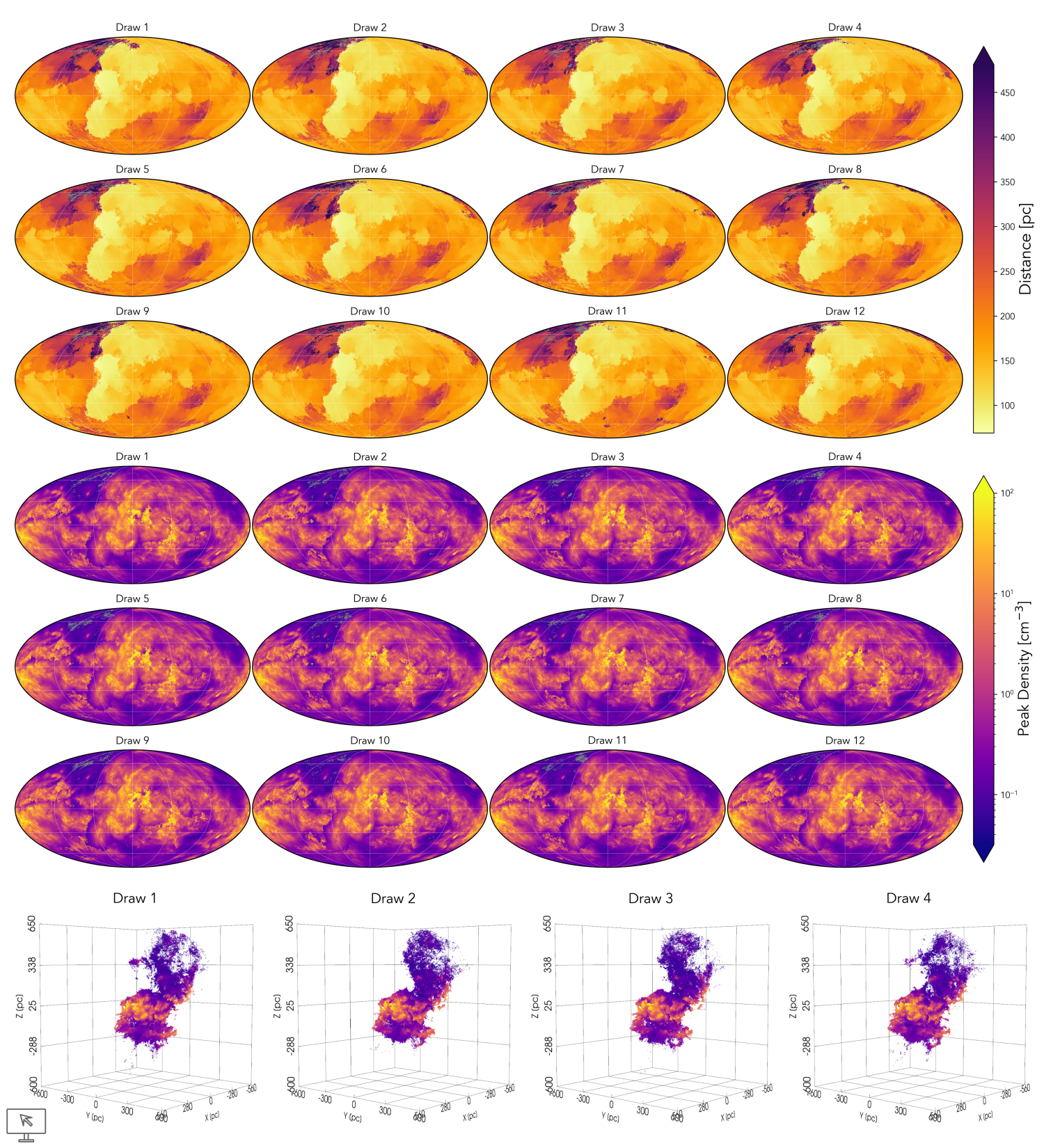}}
    \caption{\textit{Top rows:} Mollweide projections of distance to the first peak along the LOS in each of the 12 draws of the E24 map are shown.  \textit{Center rows:} As top, but for peak density.  \textit{Bottom row:} 3D views of models derived from draws 1, 2, 3, \& 4, colored by peak density.  \underline{Interactive 3D figure} showing all 12 draws: \url{https://theo-oneill.github.io/localbubble/draws/}.  To animate the figure, click "Play loop" in the interactive control panel.}
    \label{fig:multi_draws}
\end{figure*}

\begin{figure*}
    \centering
    \href{https://theo-oneill.github.io/localbubble/sigma_distance/}{\includegraphics[width=\textwidth]{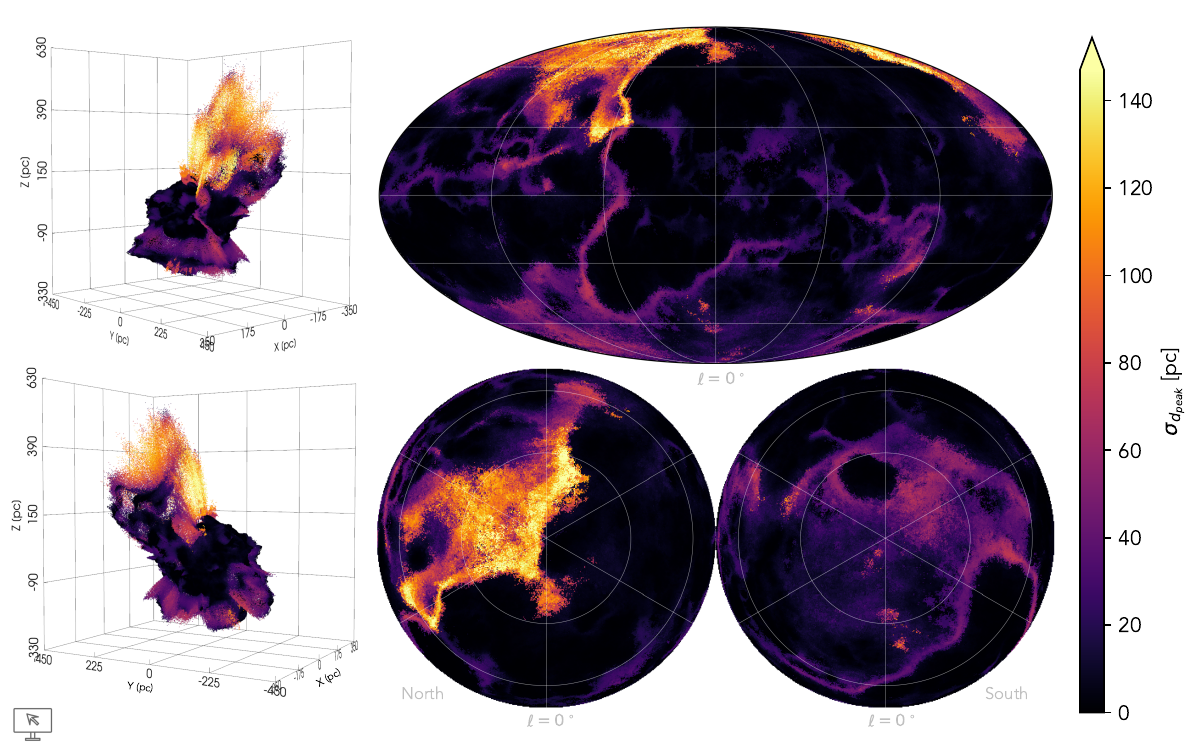}}
    \caption{Uncertainty on distance to the \LB's shell, $\sigma_{d_{peak}}$, derived from the 12 draws of the E24 map, is shown in 3D and in 2D projections.  Distance in 3D projection is the mean position in all 12 draws.  \underline{Interactive 3D figure}: \url{https://theo-oneill.github.io/localbubble/sigma_distance/}.}
    \label{fig:sig_distance_peak}
\end{figure*}

To estimate statistical uncertainties on the properties of the \LB's shell, we apply our peak finding process to each of the 12 draws of the E24 map. The individual draws are generally noisier along the LOS than the posterior mean; we thereform perform our optimal \Pmin selection procedure (described in Appendix \ref{ap:peakfind_params}) for each draw, and select a \Pmin to be used on all 12 draws as the knee of their mean \Pmin vs. outlier \Npeak fraction curve.  This yields an optimal \Pmin=$1.26 \times 10^{-5}$ ZGR23/pc to be used for peak finding in all draws.  

We define statistical uncertainties on each derived property of the \LB's shell as the standard deviation of the equivalent properties in the 12 draws.  We expect this procedure should provide a reasonable estimate of statistical uncertainties introduced by the Gaussian Process-based construction of the E24 map; however, it does not account for systematic uncertainties involved in the creation of the map (see E24 for a discussion of these other uncertainties and caveats).  Our uncertainty estimates along individual LOS also do not include systematic uncertainties introduced by the assumptions used to convert from measured differential extinction to derived quantities like extinction, volume density, and mass.

As an example of the results of this procedure, Figure \ref{fig:sig_distance_peak} shows the uncertainty on distance to the \LB's shell, $\sigma_d$.  We find that $\sigma_d$ is inversely correlated with dust density, and that $\sigma_d$ is typically less than 10 pc  in most of the higher density regions.  $\sigma_d$ is highest in the Northern chimney region, which is unsurprising given how low density the dust is and how wide the shell appears to be in the posterior mean model in this area. This suggests that our estimate of shell thickness in this region is merely an upper limit, and that shell dust is likely contained somewhere within this wide range.  

\section{Comparison to the Pelgrims et al. (2020) Local Bubble Model}\label{ap:pelgrims}
\restartappendixnumbering

\begin{figure*}
    \centering
    \href{https://theo-oneill.github.io/localbubble/compare_pelgrims/}{\includegraphics[width=\textwidth]{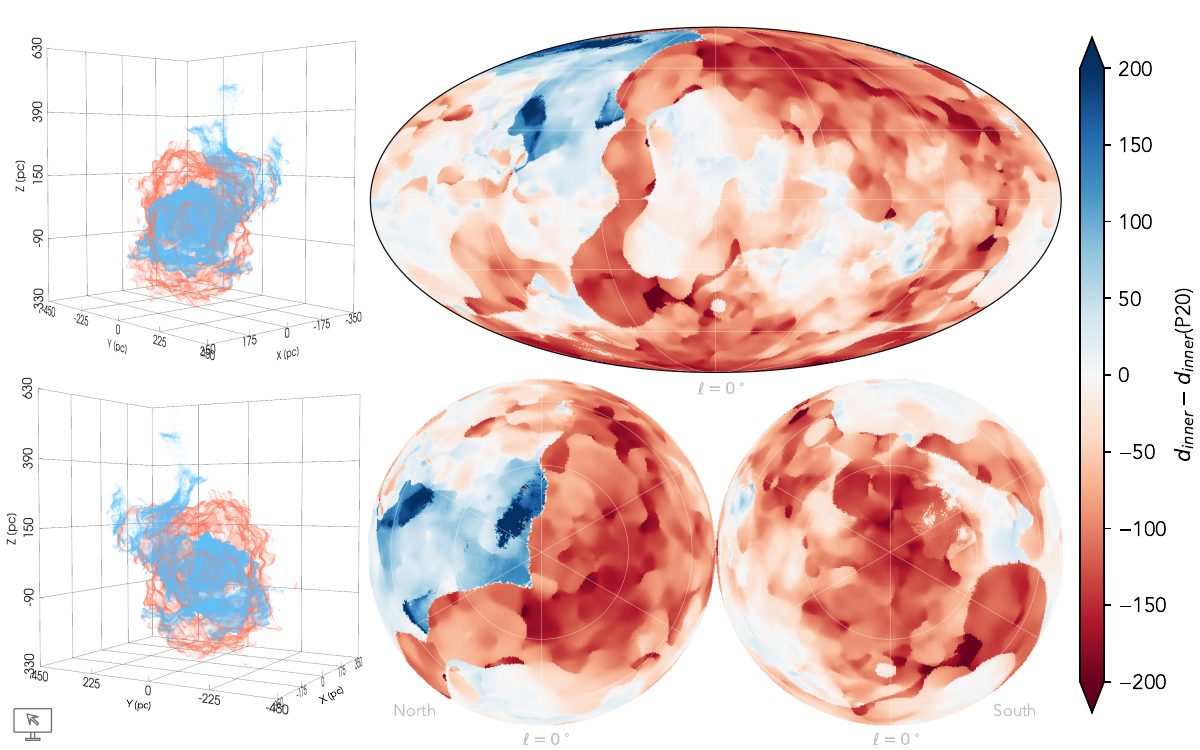}}
    \caption{Difference between distance to the \LB's inner surface in our model vs. the Pelgrims et al. (2020) model is shown in 3D and 2D projections.  \underline{Interactive 3D figure}: \url{https://theo-oneill.github.io/localbubble/compare_pelgrims/}}
    \label{fig:comp_pelgrims}
\end{figure*}

The model of the \LB that has been most widely adopted in recent years was created by \citet[][herafter P20]{PelgrimsFerriere2020} using \citet[][herafter L19]{LallementBabusiaux2019}'s 3D dust reddening map generated from \gaia DR2 and 2MASS photometry and astrometry.  P20's model of the \LB is noticeably different from the model of the \LB we present in this work, both in terms of distance to the shell and in interpretation of whether the \LB is open vs. closed.  In this appendix, we explore potential causes of these differences.

We expect the choice, construction, and resolution of the underlying 3D dust map to have the largest impact on models of the \LB.  The L19 and E24 maps generally agree in their reconstructions of the distances to high extinction dust features in the Solar neighborhood, but diverge in the intermediate-to-low extinction regime.  The L19 map covers a larger volume than the E24 map in the Galactic plane, but smaller towards Galactic north and south (extending from $|x| < 3$ kpc, $|y| < 3$ kpc, $|z| < 400$ pc), and is defined at a lower spatial resolution (a minimum of 25 pc within 1 kpc) than the E24 map.    

To build their model of the \LB, P20 sampled differential extinction $A'_v(r) = dA_v/dr$ along the LOS at distance intervals of 2.5 pc with an angular spacing of \Hpx \nside{128} (27' pixel size), and smoothed each LOS with a Gaussian kernel of $\sigma_{smooth} = 25$ pc.  They then defined the inner surface of the \LB as the location of the first inflection point in $A'_v(r)$ ($d^2 A'_v(r) / dr^2 = 0$) where the curve transitioned from convex to concave (i.e., from $d^2 A'_v(r) / dr^2 > 0$ to $d^2 A'_v(r) / dr^2 < 0$). They similarly defined the outer surface as the first inflection point after the inner surface where the curve transitioned from concave to convex.  

P20's model was created to help model the Local Bubble's magnetic field and infer the orientation of the Galactic magnetic field in the Solar neighborhood, making a smoothed model of the \LB's surface desirable. To this end, P20 performed iterative spherical harmonic expansions of the inner surface (with a given maximum expansion \lmax) to generate smoothed models.  Data products released by P20 included the unsmoothed distance to the inner surface as well as smoothed maps of the inner surface at a variety of \lmax.   \citet{ZuckerGoodman2022} used the P20 model with \lmax = 10 for their study of \LB-triggered star formation in the Solar Neighborhood.

P20's inflection point-based procedure was appropriate for defining the inner edge of the \LB in the L19 map due to the relatively large smoothing kernel employed along the LOS ($\sigma_{smooth} = 25$ pc, matching the approximate resolution of the L19 map).  We perform a variety of experiments exploring the effects of smoothing kernel size and inflection vs. prominence based methods for defining the Local Bubble's shell.  We find that the presence of a distant, low-density ``Chimney'' feature is constant under all methodological choices that we consider.  We additionally find that, for the higher resolution E24 map, we would need to overly smooth our LOS to yield inflection points that span the majority of the observable rise and decline in dust extinction preceding and following a peak.  Since many peaks in the E24 map have widths of $<$10 pc and/or fall close to the start of the dust volume probed at $d=69$ pc, employing an overly large level of smoothing causes narrow peaks to blend together and for peaks closer to the Sun to be entirely missed. We therefore prefer to define inner and outer peak edges via a simple prominence-based criterion (described in \S\ref{S:peak_find}), which we find better preserves the morphology of the unsmoothed E24 map.  

Figure \ref{fig:comp_pelgrims} compares our model of the \LB's inner surface (degraded to \nside{128} resolution) to P20's unsmoothed model in both 2D and 3D.  In low latitude regions where dust density is high, the two models are very similar.  At higher latitudes, our model generally falls at smaller distances (closer to the Sun) than P20, with the exception of the Northern Chimney feature; P20 finds closed caps to both the North and South, while we find an open Northern cap.  

\begin{figure*}
    \centering
    \includegraphics[width=\textwidth]{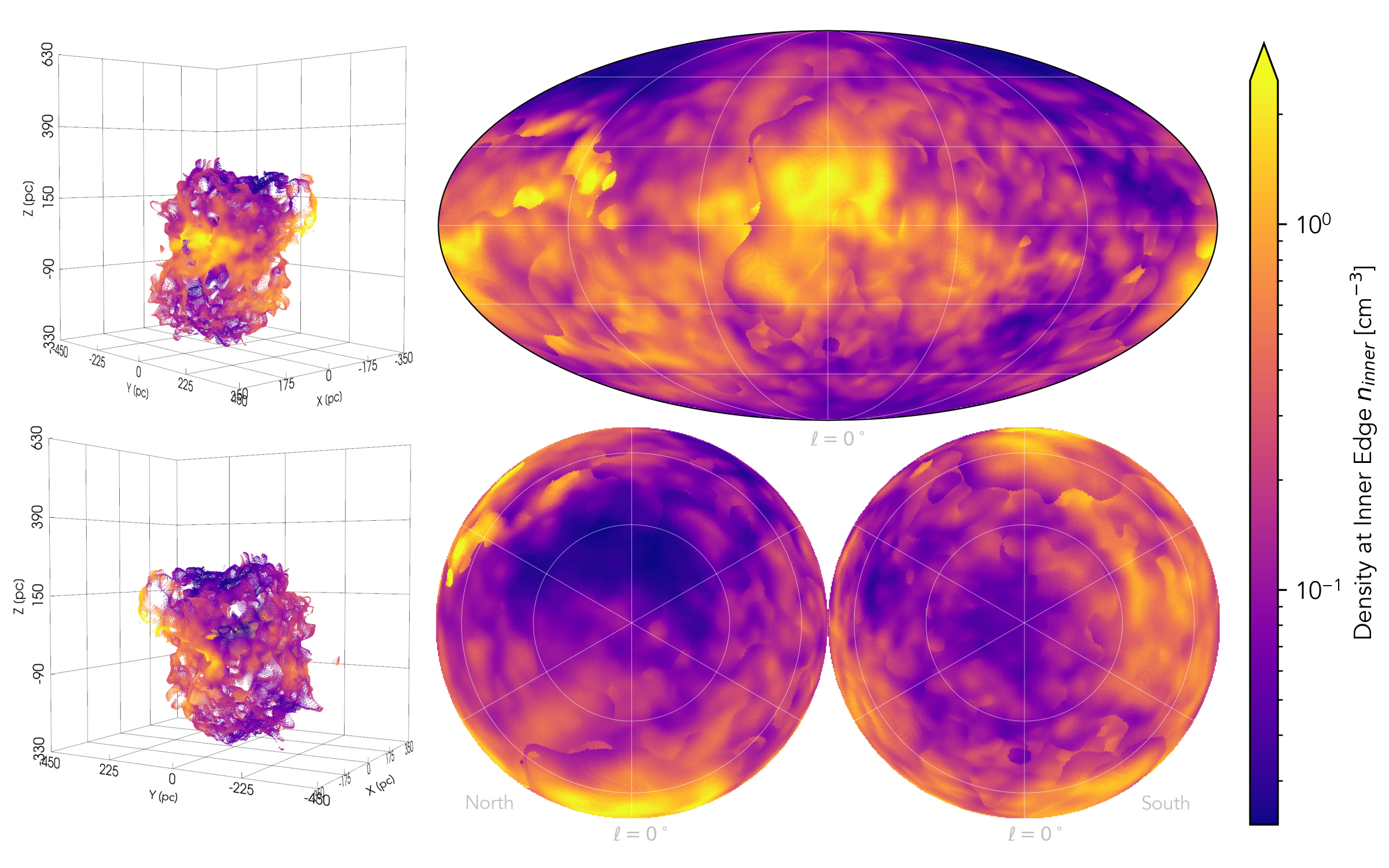}
    \caption{Volume density of dust in the \citet{LallementBabusiaux2019} dust map at the inner edge of the \LB model derived by \citet{PelgrimsFerriere2020}, in 2D and 3D projection.}
    \label{fig:pelgrims_density}
\end{figure*}

The open Northern cap in our model is characterized by a region of increased distances and decreased density dust.  To investigate the nature of the \LB's polar caps in the P20 model, we must consider both distance to and density of the shell in these regions in the L19 map.  To this end, we project the L19 cartesian dust map to a spherical grid (following P20, we use \Hpx \nside{128} angular spacing with 2.5 pc radial distance sampling), and extract the magnitude of differential dust extinction at P20's derived inner edges.  We convert differential extinction to an approximate hydrogen volume density via the method described in \S\ref{S:extinc_dens_mass}; the L19 map is defined in V-band ($\lambda = 550$ nm), which can be converted to the ZGR23 units using ZGR23's extinction curve ($A_{ZGR23} \simeq A_V / 2.658$) to yield a conversion factor to volume density of $n = 640 \ \textrm{cm}^{-3} A'_V$.  

We show the derived shell density at the inner edge of the P20 model in 2D and 3D projection in Figure \ref{fig:pelgrims_density}.  We observe that the P20 model also possesses a region of unusually low density dust over the Northern polar cap, and that the center of this low density Northern feature is aligned with \citet{LallementWelsh2003}'s reported tilt of the northern Local Chimney towards \ellb{180}{70}.  Thus, although distance estimates to the shell vary, this analysis suggests that the L19 and E24 maps both show notable asymmetries in the properties of the Northern vs. Southern cap. 

In summary, we attribute the majority of the quantitative differences between the P20 model and our model of the \LB to the difference in the spatial resolution and dust sensitivity of the underlying 3D dust maps.  Interpretations of an open vs. closed 
\LB \citep[especially in the context of the earlier work by e.g.,][]{LallementWelsh2003} can be reconciled by considering secondary properties of the Local Bubble’s shell such as shell density.  As future 3D maps of the Solar neighborhood are developed, we expect our understanding of the \LB's morphology to be further refined.

\section{Derivation of Volume Density from Differential Extinction}\label{ap:voldens}
\restartappendixnumbering

For a given band $X$ with extinction $A_X$, volume density can be derived from differential extinction in the E24 map through the following relationships.  Differential extinction at distance slice $i$ can be converted from ZGR23 units to $X$-band as,
\begin{equation}
   A'_{X, i} = m_X A'_{ZGR23, i},
\end{equation}
where the coefficient $m_X$ comes from ZGR23's published extinction curve.

Differential extinction is defined per parsec,
\begin{equation}
    A'_{X, i} = \frac{dA_{X, i}}{1 \rm{pc}},
\end{equation}
and can be converted to extinction by multiplying by the radial distance interval $dr_i$ between slice $i$ and $i+1$,
\begin{equation}
   A_{X, i} = A'_{X, i} dr_i = \frac{dA_{X, i}}{1 \rm{pc}} dr_i.
\end{equation}

By assuming the ratio of hydrogen column density to $X$-band extinction is constant,
\begin{equation}
   A_X / N_H \simeq \alpha \ \rm{mag} \ \rm{cm}^2,
\end{equation}
volume density can be derived as
\begin{equation}
   n_i = \frac{\#}{dv_i} = \frac{N_H da_i}{dv_i} = \frac{A_{X, i}}{\alpha} \frac{da_i}{dv_i},
\end{equation}
where $da_i$ is the physical area of the slice at distance $d_i$, and $dv_i$ is the volume between slice $i$ and $i+1$.

This leads to a relationship between ZGR23 differential extinction and volume density of,
\begin{equation}
   n_i = \frac{m_X}{(A_X / N_H)} \frac{dA_{ZGR23, i}}{1 \rm{pc}} \frac{dr_i da_i}{dv_i}.
\end{equation}
Since $dv_i = da_i dr_i$, this simplifies to
\begin{equation}
   n_i = \frac{m_X}{(A_X / N_H)} \frac{A'_{ZGR23, i}}{(3.086 \times 10^{18})}
\end{equation}
for $n_i$ in units of cm$^{-3}$, where the factor $3.086 \times 10^{18}$ is introduced to convert between parsecs and centimeters.

\section{Additional 3D Views of the Local Bubble}\label{ap:topdown}
\restartappendixnumbering

\begin{figure*}
    \centering
    \includegraphics[width=0.245\textwidth]{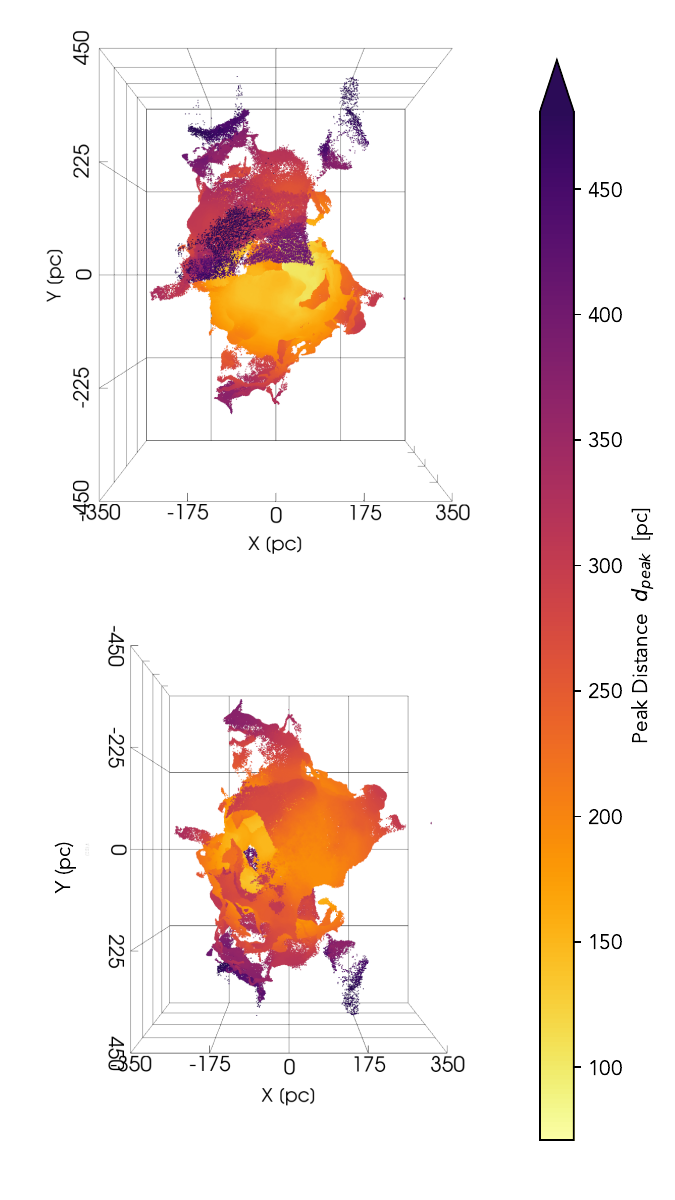}
    \includegraphics[width=0.245\textwidth]{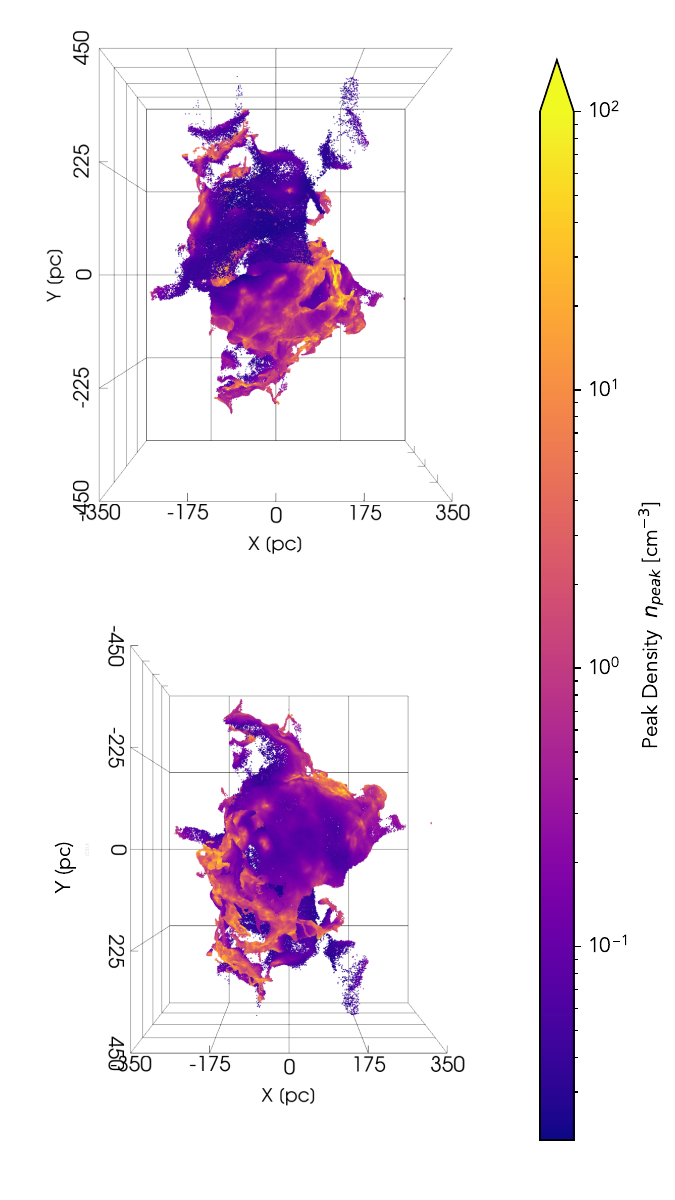}
    \includegraphics[width=0.245\textwidth]{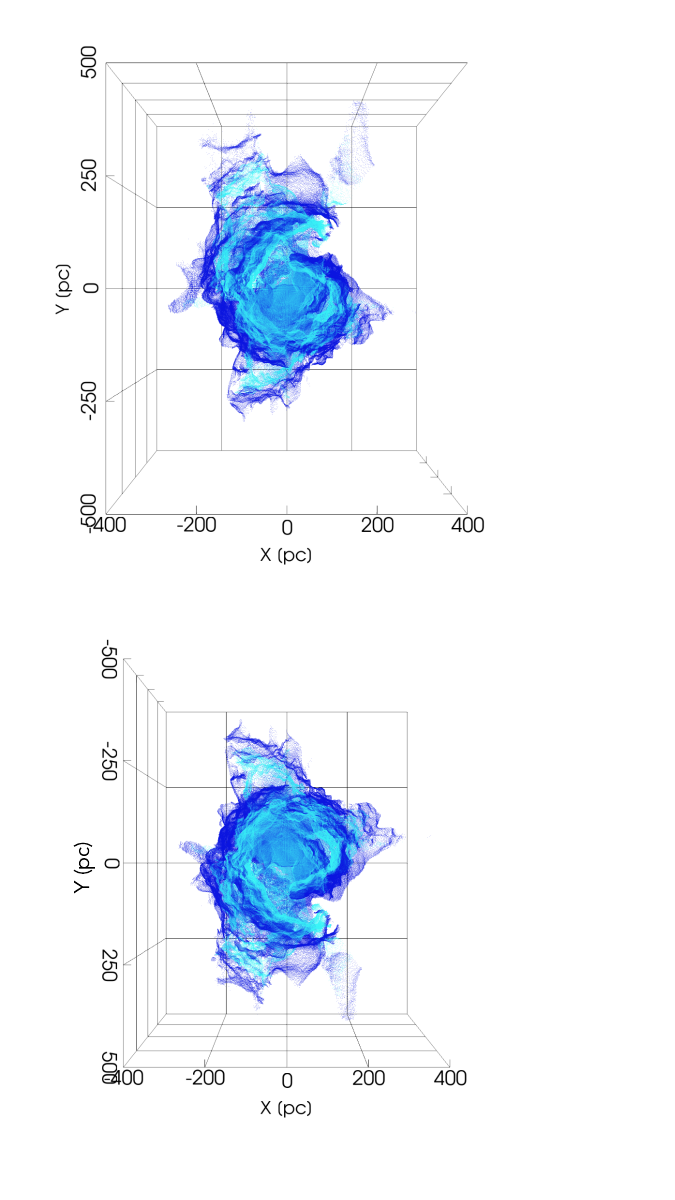}
    \includegraphics[width=0.245\textwidth]{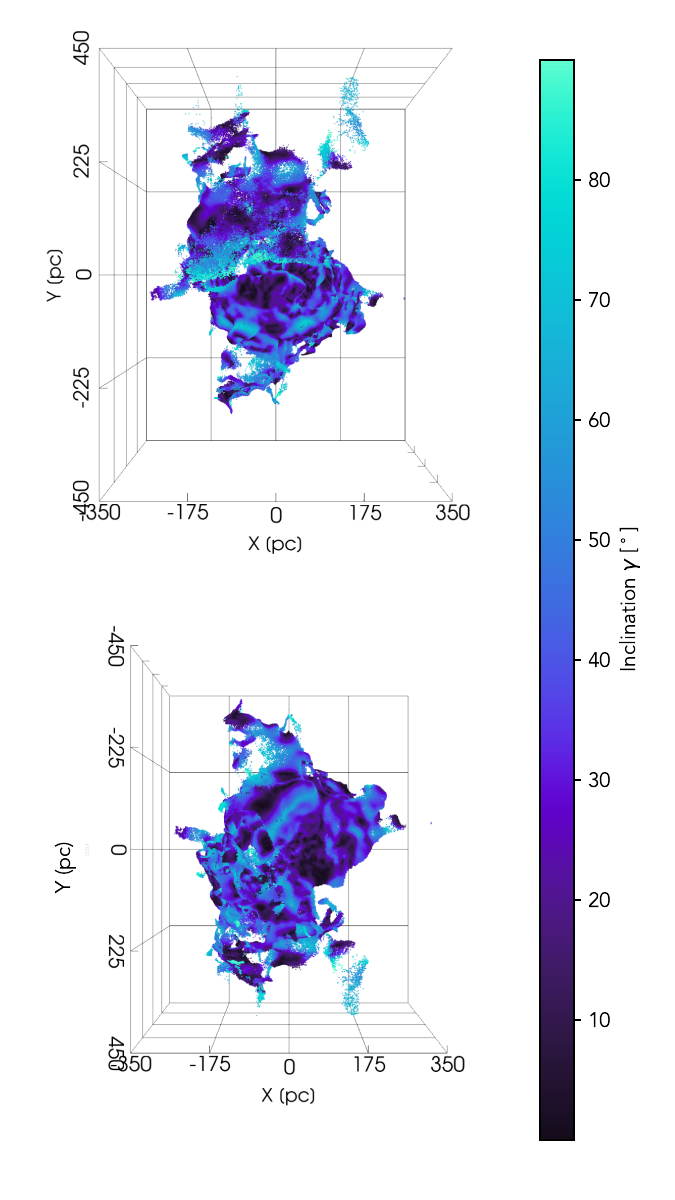}
    \caption{``Top-down'' and ``bottom-up'' 3D views of the Local Bubble properties shown in (\textit{from left}) Figures \ref{fig:distance_peak}, \ref{fig:dens_peak}, \ref{fig:thick_peak}, and \ref{fig:gamma_peak}.}
    \label{fig:topdown_props}
\end{figure*}

In Figure \ref{fig:topdown_props}, we present supplementary 3D views of the \LB properties shown in Figures \ref{fig:distance_peak}, \ref{fig:dens_peak}, \ref{fig:thick_peak}, and \ref{fig:gamma_peak} from ``above'' and ``below''.

\bibliographystyle{yahapj}
\bibliography{refs.bib,jdsbib.bib}

\end{document}